\documentclass[11pt]{JHEP3}

\pdfoutput=1

\usepackage{amssymb}
\usepackage{amsmath}
\usepackage{subeqnarray}

\allowdisplaybreaks[3]

\makeatletter
\@addtoreset{equation}{subsection}
\makeatother

\def\dalemb#1#2{{\vbox{\hrule height .#2pt
        \hbox{\vrule width.#2pt height#1pt \kern#1pt
                \vrule width.#2pt}
        \hrule height.#2pt}}}
\def\square{\mathord{\dalemb{6.8}{7}\hbox{\hskip1pt}}}

\def\cA{{\cal A}}

\newcommand{\bsea} {\begin{subeqnarray}}
\newcommand{\esea} {\end{subeqnarray}}

\newcommand{\ga}{\gamma}
\newcommand{\ba}{\beta}
\newcommand{\da}{\delta}

\newcommand{\ud}{\mathrm{d}}

\newcommand{\La}{\Lambda}

\newcommand{\sig}{\sigma}
\def\0{{\sst{(0)}}}
\def\3{{\sst{(3)}}}
\def\4{{\sst{(4)}}}
\def\5{{\sst{(5)}}}
\def\6{{\sst{(6)}}}
\def\7{{\sst{(7)}}}
\def\8{{\sst{(8)}}}
\def\n{{\sst{(n)}}}

\def\ep{\epsilon}

\def\half{{\textstyle{1\over2}}}

\let\a=\alpha \let\b=\beta \let\g=\gamma \let\d=\delta \let\e=\epsilon
\let\z=\zeta \let\h=\eta   \let\k=\kappa
\let\l=\lambda
\let\m=\mu \let\n=\nu  \let\r=\rho
\let\s=\sigma \let\t=\tau  \let\f=\phi  
\let\w=\omega

\def\nn{\nonumber} \def\bd{\begin{document}} \def\ed{\end{document}}
\def\ds{\documentstyle} \let\fr=\frac \let\bl=\bigl \let\br=\bigr
\let\Br=\Bigr \let\Bl=\Bigl
\let\bm=\bibitem
\let\na=\nabla
\let\pa=\partial \let\ov=\overline
\newcommand{\be}{\begin{equation}}
\newcommand{\ee}{\end{equation}}
\newcommand{\ch}{\mathrm{\cosh}}
\newcommand{\sh}{\mathrm{\sinh}}

\def\ba{\begin{array}}
\def\ea{\end{array}}
\def\ft#1#2{{\textstyle{{\scriptstyle #1}\over {\scriptstyle #2}}}}
\def\fft#1#2{{#1 \over #2}}
\def\del{\partial}
\def\sst#1{{\scriptscriptstyle #1}}
 \def\oneone{\rlap 1\mkern4mu{\rm l}}
\def\ie{{\it i.e.\ }}
\def\via{{\it via}}
\def\semi{{\ltimes}}
\def\str{{\rm str}}
\def\Dm{{{D_{\sst{max}}}}}
\def\vac{ \left | 0 \right \rangle }
\def\kvac{ \left | k \right \rangle }

\def\sp{\; \; \;}

\def\bol{ \left | B (p^+) \right \rangle}
\def\bo1{ \left | B^0 (p^+) \right \rangle}

\def\bolt{ \left | B (p^+) \right \rangle_{\t}}

\def\boxl{ \left | B (x^-) \right \rangle}

\def\<{ \langle }
\def\>{ \rangle }

\def\vf{\varphi}

\def\ls{{(l,0)}}
\def\lv{{(l,\pm1)}}
\def\lt{{(l,\pm2)}}

\def\lse#1{{(l_{#1},0)}}
\def\lve#1{{(l_{#1},\pm1)}}
\def\lte#1{{(l_{#1},\pm2)}}

\def\lsg#1{{5(l_{#1},0)}}
\def\lvg#1{{5(l_{#1},\pm1)}}
\def\ltg#1{{5(l_{#1},\pm2)}}

\def\lsi#1{{5{(#1,0)}}}
\def\lvi#1{{5{(#1,\pm1)}}}
\def\lti#1{{5{(#1,\pm2)}}}

\def\lsr#1{{1{(#1,0)}}}
\def\lvr#1{{1{(#1,\pm1)}}}
\def\ltr#1{{1{(#1,\pm2)}}}

\def\cD{{\cal D}}
\def\cE{{\cal E}}
\def\cF{{\cal F}}
\def\cG{{\cal G}}
\def\cH{{\cal H}}
\def\cK{{\cal K}}
\def\cO{{\cal O}}
\def\cP{{\cal P}}
\def\cQ{{\cal Q}}
\def\cR{{\cal R}}
\def\cS{{\cal S}}
\def\cT{{\cal T}}
\def\cU{{\cal U}}
\def\cV{{\cal V}}
\def\cW{{\cal W}}

\newcommand{\nono}{\nonumber}
\newcommand{\dtilde}[1]{\tilde{\tilde{#1}}}
\newcommand{\hatb}[1]{\hat{\ov{#1}}}
\newcommand{\hatt}[1]{\hat{\tilde{#1}}}
\newcommand{\emnr}{{e_\m}^{\n\r}}
\newcommand{\sub}[1]{\phantom{}_{(#1)}\phantom{}}
\newcommand{\rt}{\tilde{\r}}

\newcommand{\comment}[1]{}

\def\hna{\hat{\na}}

\newcommand{\hsp}{\hspace{0.5cm}}

\newcommand{\ho}[1]{$\, ^{#1}$}
\newcommand{\hoch}[1]{$\, ^{#1}$}
\newcommand{\bea}{\begin{eqnarray}}
\newcommand{\eea}{\end{eqnarray}}
\newcommand{\ra}{\rightarrow}
\newcommand{\lra}{\longrightarrow}
\newcommand{\Lra}{\Leftrightarrow}
\renewcommand{\ap}{\alpha^\prime}
\newcommand{\bp}{\tilde \beta^\prime}
\newcommand{\tr}{{\rm tr} }
\newcommand{\Tr}{{\rm Tr} }
\newcommand{\NP}{Nucl. Phys. }

\title{Holography for Einstein-Maxwell-dilaton theories from
generalized dimensional reduction}

\author{Blaise Gout\'eraux${}^1$, Jelena Smolic${}^2$, Milena Smolic${}^2$, Kostas Skenderis${}^{2,3}$ and Marika Taylor${}^2$\\
\\
${}^1${\it Univ Paris Diderot, Sorbonne Paris Cit\'e, APC, UMR 7164 CNRS,\\ F-75205 Paris, France} \\
\\
${}^2${\it Institute for Theoretical Physics, University of Amsterdam, \\Science Park 904, Postbus 94485, 1090 GL Amsterdam, The Netherlands} \\
\\
${}^3${\it Korteweg-de Vries Institute for Mathematics, University of Amsterdam, \\Science Park 904, 1090 GL Amsterdam, The Netherlands} \\
\\
E-mail: \email{blaise.gouteraux@apc.univ-paris7.fr}, \email{J.Smolic@uva.nl}, \email{M.Smolic@uva.nl}, \email{K.Skenderis@uva.nl}, \email{M.Taylor@uva.nl}}

\preprint{arXiv:1110.2320\\ITFA-2011-15}

\abstract{
We show that a class of Einstein-Maxwell-Dilaton (EMD) theories are
related to higher dimensional AdS-Maxwell gravity via a dimensional
reduction over compact Einstein spaces combined with continuation in the
dimension of the
compact space to non-integral values (`generalized dimensional reduction').
This relates (fairly complicated)
black hole solutions of EMD theories to simple black hole/brane
solutions of AdS-Maxwell gravity and explains their
properties. The generalized dimensional reduction is used to infer the
holographic dictionary and the hydrodynamic behavior for this class of
theories from those of AdS. As a specific example, we analyze
the case of a black brane carrying a wave whose universal sector is described
by gravity coupled to a Maxwell field and two neutral scalars. At thermal
equilibrium and finite chemical potential the two operators dual to
the bulk scalar fields acquire
expectation values characterizing the breaking of conformal
and generalized conformal invariance. We compute holographically the first
order transport coefficients (conductivity, shear and bulk viscosity)
for this system.}

\keywords{AdS-CFT Correspondence, Gauge-gravity correspondence, Black Holes, Holography and condensed matter physics (AdS/CMT)}

\begin{document}


\section{Introduction}

Recently there has been increased interest in understanding holography
for Einstein-Maxwell-dilaton theories. Such theories have the right
field content to describe holographically systems at finite charge density,
possibly in the presence of condensates, and as such they have
appeared in the holographic modeling of strongly interacting condensed
matter systems \cite{Taylor:2008tg,Gubser:2009qt,Goldstein:2009cv,Cadoni:2009xm,Chen:2010kn,Charmousis:2010zz,Lee:2010qs,Lee:2010ii,Perlmutter:2010qu,Liu:2010ka,Goldstein:2010aw,Cadoni:2011kv,Iizuka:2011hg,Gouteraux:2011ce},
see \cite{Hartnoll:2011fn} for a recent review and further references.
Since many of the relevant
solutions are not asymptotically AdS, it is not a priori clear how to set up
holography. It is the purpose of this paper to provide a holographic
dictionary for a class of such theories.

In order to set up holography one needs to understand the
asymptotic structure of the field equations, identify where the source
of the dual operator is located, evaluate the on-shell action,
determine a set of (local covariant) counterterms
and finally compute the renormalized 1-point functions in the presence
of sources (see \cite{Skenderis:2002wp,Papadimitriou:2004ap} for reviews).
This procedure has been carried out for AdS gravity (coupled to
matter fields) in
\cite{deHaro:2000xn} (see also \cite{Bianchi:2001kw,Papadimitriou:2004rz})
and for gravity coupled to a scalar field
with exponential potential in
\cite{Wiseman:2008qa,Kanitscheider:2008kd}. The latter case is
associated with the near-horizon limit of the non-conformal branes
\cite{Itzhaki:1998dd,Boonstra:1998mp}.
It was realized later in \cite{Kanitscheider:2009as} that the
two cases are actually closely related: one can obtain the results in
\cite{Kanitscheider:2008kd} from those in \cite{deHaro:2000xn}
\via  \ a `generalized dimensional reduction'.

More precisely,
it was shown in \cite{Kanitscheider:2008kd} that the $(d+1)$-dimensional
gravity-dilaton system with an exponential potential
$V(\phi) \sim \exp (-\delta \phi)$
can be obtained from AdS$_{2 \sigma +1}$ gravity
by diagonal reduction over $T^{2\sigma-d}$ torus.
This is a consistent reduction, so the structure of the
solutions of the field equations of the reduced theory
and all other results needed in order to set up
holography can be deduced from that of AdS gravity.
The results depend smoothly on $\sigma$ as long as $\sigma > d/2$.
This can be seen by inspection of the results but it is also
intuitively clear: the dimension of the torus, $(2 \sigma-d)$,
should be positive. It follows that one can use the dimensional reduction
in order to establish a holographic dictionary for this theory
when\footnote{The non-conformal branes correspond to specific rational
values of $\sigma$.} $\sigma > d/2$ (this translates into a constraint
on the slope of the potential, $\delta ^2<2/(d-1)$). Indeed, one can check
that the results established in \cite{Kanitscheider:2008kd} by
a direct analysis of the field equations etc. are reproduced exactly.
This method was also applied successfully to
probe non-conformal branes \cite{Benincasa:2009ze}:
the corresponding holographic dictionary was
obtained from the results in \cite{Karch:2005ms} in this way.

This method can be used in order to set up holography for
any theory that is related to a theory for which the holographic
dictionary is known \via\ such a 'generalized consistent reduction'.
The upstairs theory can thus be AdS gravity coupled to general
matter (scalar fields, fermions, gauge fields, form fields). The
reduction must be consistent, \ie all solutions of the lower-dimensional theory should be solutions of the higher-dimensional
theory. This is necessary in order to be able
to deduce the structure of the field
equations of the lower-dimensional theory etc. from that of the
higher-dimensional theory. The reduction
will be 'generalized' if the reduced theory depends smoothly on
the dimension of the compactification manifold (and perhaps other
such parameters) which could then be continued to be any real
number.

In this paper we will focus on a lower-dimensional theory that
contains Maxwell fields.
One way to obtain Maxwell fields is to replace
the diagonal torus reduction by a general non-diagonal reduction
and another is to have Maxwell fields already in the upstairs theory.
The diagonal torus reduction can also be replaced by a
reduction over an Einstein manifold (which is also
consistent as long as we only keep
the mode parametrizing the overall
size of the compact manifold). Such reduction produces a lower-dimensional
theory with a potential having two exponential terms.
Yet another possibility is to start
with form fields in higher dimensions. This case has been analyzed in
\cite{Gouteraux:2011ce} and it will not be discussed here.

For the applications of interest, one would like to have explicit
black hole solutions where the scalar and Maxwell fields are non-trivial.
It turns out that the theories obtained \emph{via} a generalized
dimensional reduction from a higher-dimensional AdS gravity (possibly coupled
to a Maxwell field) are the
same as the theories where non-extremal black hole solutions are
explicitly known. The asymptotics of these solutions
are often 'unconventional' as
the scalar field  blows up at the infinity. This behavior complicates the computation of
conserved charges. Our results streamline this discussion as well:
conserved charges can be computed using the holographic stress
energy tensor and the holographic conserved current. Since these
objects originate from their higher-dimensional AdS counterparts,
they satisfy all expected thermodynamic identities, just as the AdS ones, \cite{Papadimitriou:2005ii}. The  solutions themselves
(which often look complicated) originate from simpler solutions in
higher dimensions.
Moreover, one can also obtain a description of the hydrodynamic
regime from that of the higher-dimensional AdS case. This leads to
the derivation of the relevant transport coefficients and can explain
some special relations they may satisfy. For example, these considerations
explained in \cite{Kanitscheider:2009as}
that the special value of the bulk to shear viscosity ratio
for all backgrounds which asymptote to the non-conformal brane
background is due to the conformal symmetry of the higher-dimensional
theory. The same method was also used in \cite{Bigazzi:2010ku} in
order to compute transport coefficients for the Quark-Gluon Plasma
using a holographic model of QCD.

This paper is organized as follows. In the next section, we list the
cases where the Einstein-Mawell-dilaton theory can be oxidised to
a higher-dimensional AdS-Maxwell theory and we discuss their
black hole solutions, their uplift to AdS black holes and
how their charges and thermodynamics can be explained \emph{via} the
lift to higher dimensions. Then, in section 3 we specialize to
one of the relevant cases and we fully carry out the program
discussed above. Finally, in appendix A we prove that the dimensional
reductions used in this paper are consistent.

\section{Oxidation of Einstein-Maxwell-Dilaton theories} \label{section:oxi}

In this section we will consider how higher-dimensional AdS(-Maxwell) gravity reduces to Einstein-Maxwell-Dilaton (EMD) theories \emph{via} a (generalized) consistent (non-)diagonal Kaluza-Klein reduction.
We will further connect the properties of
the (EMD) black hole solutions with those of higher-dimensional black
holes, charged (asymptotically flat) black $p$-branes and boosted black branes.

The higher-dimensional action is given by
\be
S_{(2\sig+1)} = \frac{1}{16\pi G_N^{(2\sig+1)}}\int_{\mathcal M}\ud^M
x\,\ud^a y\sqrt{-g_{(2\sig+1)}}\left[R_{(2\sig+1)}- \frac14
  F^2-2\Lambda\right] \label{EMLambdaAction} \ee
The integral is over
the bulk $(2 \sigma+1)$-dimensional spacetime $\mathcal M$.
Capital latin indices
$M,N,\ldots$ run from $0$ to $d$, and denote lower,
$(d+1)$-dimensional spacetime coordinates, while lowercase latin
indices $a,b,\ldots$ will typically run from $d+1$ to $2\sig$ and
denote internal coordinates. Highercase latin indices $A,B,\ldots$
refer to the higher-dimensional spacetime coordinates and run from $0$
to $2\sig$.  Lowercase greek indices $\mu,\nu,\ldots$ are higher-dimensional indices and run from $0$ to $2\s-1$, while lowercase latin indices $i,j,\ldots$ are lower-dimensional boundary indices and run from $0$ to $d-1$. The Maxwell terms with straight latin uppercase originate from higher dimensions, $A=A_A\ud x^A$, with field strength \be
F=\half F_{AB}\ud x^A\wedge\ud x^B = \ud A, \qquad
F_{AB}=2\partial_{\left[A\right.}A_{\left.B\right]}, \ee where the
brackets are the usual antisymmetry operation. The Maxwell term is
invariant upon taking the Hodge dual $\star F$ of the field strength,
which allows to generate magnetic solutions from electric solutions
and vice-versa.

In what follows we will be interested in
Kaluza-Klein reductions on a $(2\sig-d)$-dimensional internal space
$\mathbf X_{(2\sig-d)}$ times a $(d+1)$-dimensional manifold $\mathcal M_{(d+1)}$; details of the compactifications are given in Appendix \ref{Appendix}.
The reductions of interest are over Einstein manifolds. Recall that a $p$-dimensional Einstein manifold $\mathbf X_{(p)}$ satisfies
\be
    R^{(p)}_{ab} = (p-1) \lambda_{(p)}g_{ab}\,.
    \label{EinsteinSpace}
\ee
We will denote the metric of  $\mathbf X_{(p)}$ by $\ud s^2= \ud X_{(p)}^2$ and its volume by $V_{(p)}$.
When the Einstein manifold is homogeneous, this implies that the Riemann tensor is
\be
    R^{(p)}_{abcd}= \lambda_{(p)}\left(g_{ac}g_{bd}-g_{ad}g_{bc}\right),
    \label{ConstantCurvatureSpace}
\ee
where $\lambda_{(p)}$ is then normalized to $\pm1,0$.
We also define the Anti-de Sitter radius in $2\sig+1$ dimensions as:
\be
    -2\La\ell^2_{(2\s+1)} = 2\sig(2\sig-1)\,.\label{AdSradius}
\ee

\subsection{Diagonal reduction to Einstein-Dilaton theories\label{section:NED}}

Let us start from the AdS-Einstein action in $2\sig+1$ dimensions:
\be
    S_{(2\sig+1)}=\frac1{16\pi G_N^{(2\sig+1)}}\int_{\mathcal M} \ud^{2\sig+1}x\,\sqrt{-g_{(2\sig+1)}}\left [R-2 \La \right].
    \label{EinsteinAdSAction}
\ee
We show in appendix \ref{Appendix} that the reduction ansatz
\be
    \ud s^2_{(2\sigma+1)}=e^{-\da_1\phi}\ud s^2_{(d+1)} + e^{\frac{\phi}{\da_1}\left(\d_c^2-\da_1^2\right)}\ud X^2_{(2\sigma-d)}
    \label{KKDiagonal}
\ee
with
\be
    \da_1^2=\frac{2(2\sig-d)}{(d-1)(2\sig-1)}\leq \delta_c^2 \equiv \frac{2}{(d-1)} \quad \Longleftrightarrow \quad 2\sig = \frac{d-\frac{d-1}2\da_1^2}{1-\frac{d-1}2\da_1^2}\geq0\,,\label{AdSDaSigRelation}
\ee
consistently reduces \eqref{EinsteinAdSAction} to a $(d+1)$-dimensional theory
with action
\be
    S_{(d+1)}=\frac1{16\pi G_N^{(d+1)}}\int_{\mathcal M} \ud^{d+1}x\,\sqrt{-g_{(d+1)}}
\left [R-\half(\partial\phi)^2-2\La_1 e^{-\da_1\phi}-2\La_2 e^{-\da_2\phi} \right ],
    \label{ED2LiouvilleAction}
\ee
where
\be \label{ident}
\La=\La_1\,,\qquad  R_{(2\sigma-d)}=-2\La_2\,,\qquad \da_2=\frac{\d_c^2}{\da_1}\geq \da_c.
\ee
Note that consistency of the reduction requires that $\mathbf X_{(2\sig-d)}$
is an Einstein manifold.

The Einstein-Dilaton theory we obtain has a scalar potential comprising
two exponential terms, whose origin are respectively the
higher-dimensional cosmological constant, $\La$, and the curvature $R_{(2\sigma-d)}$ of
the internal space. Note that in this case, the
slope of the exponential $\da_1$ is restricted to the interval
$[0,\da_c]$ so that the number of reduced dimensions $2\sigma-d$
remains positive. Consequently, the slope of the exponential $\da_2$
is restricted to the complementary interval $[\da_c,+\infty]$.

Since the action \eqref{ED2LiouvilleAction} is invariant under the exchange $\{\La_1,\da_1\} \leftrightarrow \{\La_2,\da_2\}$,
it may also be obtained from the Einstein-AdS action \eqref{EinsteinAdSAction} by the alternate reduction ansatz,
\be \label{KKDiagonal2}
    \ud s^2_{(2\sigma+1)} = e^{-\frac{\d_c^2}{\da_1}\phi}\ud s^2_{(d+1)}+e^{(\d^2_1-\d_c^2)\frac{\phi}{\d_1}}\ud X_{(2\sig-d)}^2\,,
\ee
with
\be
    \da^2_c<\da_1^2 = \frac{2(2\sigma-1)}{(d-1)(2\sigma-d)}<\da^2_{max}\equiv \frac{2d}{(d-1)} \quad \Longleftrightarrow \quad +\infty>2\sigma-d>1\,, \label{BpBDaSigRelation}
\ee
\be
 \La=\La_2\,, \qquad  R_{(2\sig-d)}=-2\Lambda_1\,,\qquad \da_2 = \frac{\d_c^2}{\da_1}\leq \d_c^2\,,
\ee
Note that the upper bound on the value of $\da_1$ corresponds to the internal space being one-dimensional, \emph{i.e.} $R_{(1)}=-2\Lambda_1=0$.

Since $\sigma$ is related to the dimension of the higher-dimensional
theory \emph{via} $D=2 \sigma +1$, it should be a (half) integer. However,
after reduction $\sigma$ enters algebraically as a parameter in the
$(d+1)$-dimensional action, so one may analytically continue its value to any
real number \cite{Kanitscheider:2009as},
modulo restrictions that arise from the requirement that
the lower-dimensional theory is well-behaved (kinetic terms should be
positive definite etc.). This generates
the continuous family of theories \eqref{ED2LiouvilleAction} labeled
by a real parameter $\da_1$, related to a higher-dimensional AdS-Mawxell theory \emph{via} generalized dimensional
reduction.

Analyzing the equations of motion derived from action
\eqref{ED2LiouvilleAction}, one can show that there exist analytic
black hole solutions precisely when the theory is related to a higher-dimensional AdS theory, namely when $\da_2=\d_c^2/\da_1$.
The solution is given by, \cite{Chan:1995fr,Cai:1996eg,Cai:1997ii,Charmousis:2001nq},
\bea
    \ud  s^2_{(d+1)} &=&-V(r)\ud t^2+\frac{e^{\da_1\phi}\ud r^2}{V(r)}+r^2\ud X^2_{(d-1)}\,, \label{EDBH2Liouville}\\
    V(r)&=&\left(\frac r\ell\right)^2+\frac{(d-2)\lambda_{(d-1)}r^{(d-1)\da_1^2}}{\left(1-\frac{d-1}{2}\da_1^2\right)\left(d-2+\frac{d-1}{2}\da_1^2\right)}-mr^{2-d+\frac{d-1}2\da_1^2}\,,\label{PotEDBH2Liouville}\\
    e^\phi&=&r^{(d-1)\da_1}\\
    -2 \Lambda_2&=&\frac{(d-1)^2(d-2)\da_1^2\lambda_{(d-1)}}{2\left(1-\frac{d-1}2\da_1^2\right)} \label{LakRel1}\\
    -2\La_1\ell^2&=&(d-1)\left(d-\frac{d-1}{2}\da_1^2\right)\,. \label{EDAdSradius}
\eea
The solution
has a curvature singularity at $r=0$ and an event horizon wherever $V(r_+)=0$. One may set $\lambda_{(d-1)}=0$ in
the above expression and obtain the generic neutral planar black hole solution, whose horizon has topology $\mathbf R^{d-1}$.
In this case the scalar potential reduces to a single exponential.
At first sight it may seem that the solution is singular in the limit $\da^2\to\da^2_c$, but a scaling limit can be taken if one
simultaneously sends $\lambda_{(p-1)}\to0$ while keeping the ratio of the two previous quantities fixed.

The action \eqref{ED2LiouvilleAction} is symmetric under the exchange
\be
    \Lambda_1 \longleftrightarrow \Lambda_2\,,\qquad \d_1 \longleftrightarrow \d_2\,, \label{EDDuality}
\ee
which allows to generate a solution dual to \eqref{EDBH2Liouville}.

\paragraph{Thermodynamics.}

The thermodynamics of these black holes may be calculated by
computing the Euclidean on-shell action and taking appropriate derivatives of the thermodynamic potential with respect to the thermodynamic variables, \emph{e.g.}:
\be
    S[T]=-\frac{\ud F[T]}{\ud T}\,,\quad M[T]=F[T]-T\frac{\ud F[T]}{\ud T}\,,\quad F[T]=M[T]-T\,S[T].
\ee
in the canonical ensemble.

These results may be put on firmer footing. As we discuss in detail in section \ref{Gen_Hol}, the generalized dimensional reduction
leads to a holographic stress energy tensor, which may be used to compute the various thermodynamic quantities (such as the mass of the black hole).
Either way, these computations lead to the results:
\bea
     T_{(d+1)} &=&  \frac{r_+^{1-\frac{d-1}2\d_1^2}}{4\pi}\left[\left(d-(d-1)\frac{\d_1^2}2\right)\ell^{-2}+\frac{(d-2)\l_{(d-1)}}{\left(1-\frac{d-1}{2}\d_1^2\right)}r_+^{(d-1)\d^2-2}\right], \label{TED}\\
    S_{(d+1)}&=&\frac{V_{(d-1)}}{4G_N^{(d+1)}}r_+^{(d-1)}, \label{SED}\\
    M_{(d+1)}&=&\frac{V_{(d-1)}(d-1)r_+^{d-\frac{d-1}2\d_1^2}}{16\pi G_N^{(d+1)}}\left[\ell^{-2}+\frac{(d-2)\l_{(d-1)}r_+^{(d-1)\d_1^2-2}}{\left(1-\frac{d-1}2\d_1^2\right)\left(d-2+\frac{d-1}2\d_1^2\right)}\right]. \label{MED}
\eea
$V_{(d-1)}$ stands for the volume of $\mathbf X_{(d-1)}$. One may check that the first law holds:
\be
    \ud M=T\ud S\quad \Leftrightarrow \quad \ud F = -S\ud T\,,
\ee
and then examine global and local thermodynamical equilibrium by computing the free energy and the heat capacity:
\bea
    F_{(d+1)}&=&\frac{V_{(d-1)}\left(1-\frac{d-1}2\d_1^2\right)r_+^{d-\frac{d-1}2\d_1^2}}{16\pi G_N^{(d-1)}}\left[-\ell^2+\frac{(d-2)\l_{(d-1)}r_+^{(d-1)\d_1^2-2}}{\left(1-\frac{d-1}2\d_1^2\right)\left(d-2+\frac{d-1}2\d_1^2\right)}\right],\label{GED}\\
    C&=&\left.T\frac{\ud S}{\ud T}\right.\,.
\eea
We will shortly see that these results descend from higher dimensions.

\bigskip

The Einstein dilaton action with two exponential terms in its potential was already recognized as descending from a higher-dimensional Einstein action
by the authors of \cite{Mignemi:1988qc, Wiltshire:1990ah}, but they did not consider the neutral, dilatonic solution
\ref{EDBH2Liouville} from the point of view of the higher-dimensional theory, while the authors of
\cite{Cai:2004iy} only considered the oxidation of \eqref{EDBH2Liouville} with a single potential turned on, and also not in the context of 'generalized' reductions.

From the previous considerations and given a specific exponential scalar potential
\be
    V(\phi)=V_0e^{-\delta_1 \phi},
\ee
we now understand it can descend from a higher-dimensional theory in two ways, as explained in more details in Appendix \ref{Appendix:DiagonalKK}. If $\delta_1^2\leq \delta_c^2$, $V_0$ may be identified with a higher-dimensional cosmological constant, while if $\delta^2_c\leq\delta_1^2<\delta_{max}^2$, it may be identified with the curvature of the internal space $\mathbf X_{(2\sigma-d)}$ over which the reduction is performed. $\delta_1^2=\delta^2_c$ corresponds to an infinite number of dimensions, while $\delta_1^2=\delta_{max}^2$ to a single one so that the reduction over $\mathbf X_{(1)}$ does not generate a potential.

In the case where the theory has a single potential and for planar black holes, it was shown in \cite{Gursoy:2007er,Gursoy:2008za,Charmousis:2010zz} that for $\da_1^2\leq\da^2_c$, the spectrum of fluctuations was continuous and gapless, while for $\delta^2_c>\delta_1^2$, it was discrete with a gap. Moreover, this gives a recipe for generating first-order phase transitions in EMD theories, while imposing a planar boundary: by considering a potential with two exponentials, with slopes verifying $\d_1\leq\d_c$ and $\d_2>\d_c$, the former should act as a cosmological constant, the latter as horizon curvature. This matches with the intuition from the KK reduction and was exhibited in $d=4$ in \cite{Gursoy:2008za}.

We will now consider the uplift of the solution \eqref{EDBH2Liouville}-\eqref{EDAdSradius} for the two different ranges of $\delta_1$.
As we will see they originate from different higher-dimensional spacetimes in the two respective cases.

\paragraph{Oxidation for $\da_1^2\leq\da^2_c\,$: \label{BHOxidation}}

In this case we should use the oxidation ansatz \eqref{KKDiagonal} with $\da_1$ given by \eqref{AdSDaSigRelation}.
The uplift of the solution \eqref{EDBH2Liouville}-\eqref{EDAdSradius} is then
\bea
    \ud s^2_{(2\sigma+1)}&=&-\frac{f(\r)}{\r\ell^2_{(2\s+1)}}\ud \tau^2 + \frac{\ell^2_{(2\s+1)}\ud \rho^2}{4\r^2f(\r)}+\rho^{-1}\left(\ud X^2_{(d-1)}+\ud X^2_{(2\sig-d)}\right),\label{AdSBHKK}\\
    f(\r)&=&1+\ell^{2}_{(2\s+1)}\left(\lambda_{(2\sig-1)}\r-m\rho^{\s} \right)\,.
\eea
To obtain this result, we have used the change of coordinates:
\be
    r^{1-\frac{d-1}2\da_1^2}= r^{\frac{d-1}{2\sig-1}}=\rho^{-\frac12} \,,\qquad \tau=\frac{2\sig-1}{d-1}t\,,\label{AdSCoordChangeNeutral}
\ee
and normalised the curvature on the horizon as
\be \label{HigherDimConstraintCurvatures}
    (2\sig-2) \lambda{}_{(2\sig-1)}=(d-2) \lambda_{(d-1)}=(2\sig-d-1) \lambda_{(2\sig-d)} = \frac{ -2\La_2}{2\sig-d}\,.
\ee
The relation \eqref{LakRel1} can now be understood from the higher-dimensional perspective as necessary
for the space $\mathbf X_{(d-1)}\times \mathbf X_{(2\sig-d)}$ to solve the higher-dimensional Einstein equations.
The uplifted spacetime is then simply the Schwarzschild-AdS$_{(2\sig+1)}$ black hole,
where the horizon topology is not $\mathbf X_{(2\sig-1)}$ but the product space $\mathbf X_{(d-1)}\times \mathbf X_{(2\sig-d)}$. Their
normalised curvatures $\lambda_{(2\sig-d)}$, $\lambda_{(d-1)}$ must satisfy \eqref{HigherDimConstraintCurvatures}: as a consequence, only one of the $\lambda$ may generically be set to $\pm1,0$, except if $2\sig-d=d-1$ (identical compact spaces). For spheres, this means they cannot have the same radius.
As remarked in the previous section, the horizon curvature can be set to zero, in which case $\La_2=0$ and the AdS planar black hole is recovered. In higher-dimensional Einstein gravity,  the requirement that the horizon is homogeneous is relaxed to being simply Einstein: this is essential to our ability to carry out the generalized reduction of higher-dimensional solutions in order to generate lower-dimensional ones.

The thermodynamics of asymptotically (locally) AdS spaces can be
worked out using standard holographic technology
\cite{Papadimitriou:2005ii}. In particular,
the holographic stress-energy tensor can be computed using
\eqref{conf_dict}: the knowledge of the $\s$-th term in the
Fefferman-Graham expansion is enough when
the spacetime has a flat boundary. For curved boundaries, one needs
to include additional terms (see \cite{Skenderis:2000in} for a
review). Let us
work out in more detail the case for $\s=2$, that is a Schwarzschild
black hole in AdS$_5$ with boundary $\mathbf S^1\times \mathbf
S^1\times\mathbf S^2$. The general formula reads \cite{deHaro:2000xn}:
\be
    \langle T_{\mu\nu}\rangle_{2\s} = \frac{4V_{(3)}}{16\pi G_N^{(5)}\ell_{(5)}^2}\left[g_{(4)\mu\nu}-\frac{g_{(2)\mu\nu}}8\left((\mathrm{Tr}g_{(2)})^2-\mathrm{Tr}(g_{(2)}^2)\right)-\frac12g_{(2)\mu\nu}^2+\frac{\mathrm{Tr}g_{(2)}}4g_{(2)\mu\nu}\right]
\ee
while from \eqref{AdSBHKK}
\bea
    g_{(0)}&=&\mathrm{Diag}\left(-1,1,1,\sin^2\theta\right),\\
    g_{(2)}&=&\frac{\ell_{(5)}^2\l_{(3)}}{2}\mathrm{Diag}\left(1,1,1,\sin^2\theta\right),\\
    g_{(4)}&=&\frac{\ell_{(5)}^4\l_{(3)}^2}{4}g_{(0)}+\frac{\ell_{(5)}^2m}{4}\mathrm{Diag}\left(3,1,1,\sin^2\theta\right),
\eea
so that
\be
    \langle T_{\mu\nu}\rangle_{2\s} = \frac{V_{(3)}m}{16\pi G_N^{(5)}}\left[4\d_{\mu0}\d_{\nu0}+\h_{\mu\nu}\right].
\ee
Generalizing to arbitrary dimension, one finds for the spacetime \eqref{AdSBHKK}:
\bea
    T_{(2\s+1)}&=&\frac{1}{4\pi\r_+^{\frac12}}\left[\frac{2\s}{\ell_{(2\s+1)}^2}+(2\s-2)\l_{(2\s-1)}\r_+\right] \\
    S_{(2\s+1)}&=&\frac{V_{(2\s-1)}}{4G_N^{(2\s+1)}}\r_+^{\frac12-\s} \\
    M_{(2\s+1)}&=& \frac{V_{(2\s-1)}}{16\pi G_N^{(2\s+1)}}(2\s-1)m\\
    F_{(2\s+1)}&=& -\frac{V_{(2\s-1)}\r_+^{-\s}}{16\pi G_N^{(2\s+1)}}\left[\ell^2_{(2\s+1)}-\l_{(2\s-1)}\r_+\right]
\eea
which coincide with the expressions in \cite{Chamblin:1999hg,Chamblin:1999tk}.

We can now check the validity of the thermodynamic formul\ae\ \eqref{TED}-\eqref{GED}, taking care of including constant factors due to the change of coordinates \eqref{AdSCoordChangeNeutral}. Indeed, inspecting the reduction Ansatz and the definition of the thermodynamic potential from the on-shell Euclidean action, one may show that
\bea
    \b_{(2\s+1)}=\frac{2\sig-1}{d-1}\b_{(d+1)}\,,& \quad &S_{(2\s+1)}=S_{(d+1)}\\
    \b_{(2\s+1)}F_{(2\s+1)}=\b_{(d+1)}F_{(d+1)}\,,& \quad &\b_{(2\s+1)}M_{(2\s+1)}=\b_{(d+1)}M_{(d+1)}
\eea
which in turn lead to \eqref{TED}-\eqref{GED}.

\paragraph{Oxidation for $\da^2_c<\da_1^2<\da^2_{max}\,$: \label{BlackpBraneOxidation}}


In this case we use use the oxidation Ansatz \eqref{KKDiagonal2} and $\da_1$ is given by \eqref{BpBDaSigRelation}.
The uplift of \eqref{EDBH2Liouville}-\eqref{EDAdSradius} becomes:
\bea
    \ud s^2_{(2\sigma+1)} &=&-\frac{f(\rho)}{\ell_{(2\sigma-d+2)}^{2}\r}\ud\tau^2+\frac{\ell_{(2\sigma-d+2)}^{2}\ud\rho^2}{4\r^2f(\rho)}+\rho^{-1}\ud X_{(2\sig-d)}^2+\ud X_{(d-1)}^2\,,\label{AdSBlackpBrane}\\
    f(\r)&=& 1+\ell_{(2\sigma-d+2)}^{2}\left(\lambda_{(2\sig-d)}\r-m\rho^{\frac12(2\sig-d+1)}\right)\,.
\eea
with the change of coordinates
\be
    r^{\frac{d-1}2\da_1^2-1}=r^{\frac{d-1}{2\sig-d}}=\rho^{-\frac12}\,, \qquad \tau=\frac{2\sig-d}{d-1}t\,, \label{BpBChangeCoordNeutral}
\ee
and
\be
    -2\La_2=-(d-2)(2\sig-1) \lambda_{(d-1)}=\frac{(2\sigma-1)(2\sigma-d+1)}{\ell_{(2\sigma-d+2)}^2}\,, \quad -2\La_1=(2\sig-d)(2\sig-d-1) \lambda_{(2\sig-d)}\,. \label{BpBScalarConditionNeutral}
\ee
The former stems from \eqref{LakRel1}, the latter from having exchanged the roles of $\La_1$ and $\La_2$ in the oxidation.
The solution describes an AdS black hole in
$(2\sig-d+2)$ dimensions times a $(d-1)$-dimensional hyperbolic
plane\footnote{A well-known way of making the curvature of the brane
  worldvolume positive is to include a $(d+1)$-field strength in the
  action.}, with topology AdS$_{2\sig-d+2}\times \mathbf
X_{(d-1)}$. Note that if $\lambda_{(d-1)}=0$, we also need to set
$\La_2=0$ and the solution \eqref{AdSBlackpBrane} becomes the familiar
neutral black $(d-1)$-brane, where one adds $(d-1)$ flat directions to
the Schwarzschild metric.

As in the previous subsection, we can recover the appropriate $2\s+1$ behaviours for the thermodynamics of the black $(d-1)$-brane:
\bea
    T_{(2\s+1)}&=&\frac{1}{4\pi\r_+^{\frac12}}\left[\frac{(2\s-d+1)}{\ell_{(2\s+d+2)}^2}+(2\s-d-1)\l_{(2\s-d)}\r_+\right] \\
    S_{(2\s+1)}&=&\frac{V_{(2\s-1)}}{4G_N^{(2\s+1)}}\r_+^{-\frac12(2\s-d)} \\
    M_{(2\s+1)}&=& \frac{V_{(2\s-1)}}{16\pi G_N^{(2\s+1)}}(2\s-d)m\\
    F_{(2\s+1)}&=& -\frac{V_{(2\s-1)}\r_+^{-\frac12(2\s-d+1)}}{16\pi G_N^{(2\s+1)}}\left[\ell^2_{(2\s+d+2)}-\l_{(2\s-d)}\r_+\right].
\eea

The solution dual to \eqref{EDBH2Liouville} under the exchange \eqref{EDDuality} uplifts to \eqref{AdSBlackpBrane} if $\d_1^2<\d_c^2$, and to \eqref{AdSBHKK} if $\d_c^2<\d_1^2<\d_{max}^2$.

\subsection{Diagonal reduction to Einstein-Maxwell-Dilaton theories \label{section:EMDdiagonal}}

In this subsection, we would like to determine how Einstein-Maxwell theories
\be
S_{(2\sig+1)} = \frac{1}{16\pi G_N^{(2\sig+1)}}\int_{\mathcal M}\ud^M
x\,\ud^a y\sqrt{-g_{(2\sig+1)}}\left[R_{(2\sig+1)}- \frac14
  F^2-2\Lambda\right] \label{EMLambdaAction2}
\ee
can give rise to Einstein-Maxwell-Dilaton theories
\be
    S_{(d+1)}=\frac1{16\pi G_N^{(d+1)}}\int \ud^{d+1} x \sqrt{-g_{(d+1)}}\left[R-\half\partial\phi^2-\frac14e^{\ga\phi}F^2-2\La_1 e^{-\da_1 \phi}
-2\La_2 e^{-\da_2\phi}\right].
    \label{EMD2LiouvilleAction}
\ee
 \emph{via} diagonal Kaluza-Klein reduction; that is, the lower-dimensional Maxwell field originates from a higher-dimensional one\footnote{The case where such a  Maxwell field comes from a higher-dimensional $p$-form potential has  been investigated in some details in \cite{Gouteraux:2011ce}.} .

We shall consider a reduction Ansatz
\be
    A_{A}=\left(A_{M}(x^N),0\right),
\ee
for the Maxwell field, to avoid generating axionic fields in the lower-dimensional theory.

Using the Ansatz \eqref{KKDiagonal} for the metric, \eqref{EMLambdaAction2} reduces consistently to \eqref{EMD2LiouvilleAction} with
\be
    \g=\d_1<\d_c\,,\quad \La_1=\La\,,\quad \d_c<\d_2=\frac2{(d-1)\d_1}\,, \quad -2\La_2=R_{(2\s-d)}\,, \label{EMDdiagred1}
\ee
while using the Ansatz \eqref{KKDiagonal2} yields
\be
    \g=\d_2=\frac2{(d-1)\d_1}<\d_c\,,\quad \La_2=\La\,,\quad \d_c<\d_1\,, \quad -2\La_1=R_{(2\s-d)}\,. \label{EMDdiagred2}
\ee
Note that the introduction of the gauge field breaks the duality \eqref{EDDuality}: in the theory \eqref{EMDdiagred1}, exchanging $\d_1\leftrightarrow\d_2$ and $\Lambda_1\leftrightarrow\Lambda_2$ does not map back to \eqref{EMDdiagred1} but to \eqref{EMDdiagred2}, because $\g$ is mapped to $\d_2$. This means a single solution of \eqref{EMD2LiouvilleAction} may not be uplifted to two different solutions of  \eqref{EMLambdaAction2} as in section \ref{section:NED}. In both reduction schemes, $\g<\d_c\,$.

\subsubsection{Solution with two exponential-potential}

The neutral black hole solution \eqref{EDBH2Liouville} can be generalized to an already known charged solution if one sets $\ga=\da_1$
and $\da_2=2/(d-1)\da_1$. The solution becomes, see \cite{Chan:1995fr,Cai:1996eg,Cai:1997ii,Charmousis:2009xr}:
\bea
    \ud  s^2_{(d+1)} &=&-V(r)
\ud t^2+\frac{e^{\da_1\phi}\ud r^2}{V(r)}+r^2\ud X^2_{(d-1)}\,,
\label{EMDBH2Liouville}\\
    V(r)&=&\left(\frac r\ell\right)^2+\frac{(d-2)\lambda_{(d-1)}
r^{(d-1)\da_1^2}}{\left(1-\frac{d-1}{2}\da_1^2\right)
\left(d-2+\frac{d-1}{2}\da_1^2\right)}-mr^{2-d+\frac{d-1}2\da_1^2}+q^2r^{-2(d-2)},\label{PotEMDBH2Liouville}\\
    e^\phi&=&r^{(d-1)\da_1}\,,\\
    A&=&-\sqrt{\frac{2(d-1)}{d-2+\frac{d-1}2\da_1^2}}qr^{-(d-2)
-\frac{d-1}2\da_1^2}\ud t\,,\label{GaugeEMDBH2Liouville}\\
    -2\Lambda_2&=&\frac{(d-1)^2(d-2)\da_1^2 \lambda_{(d-1)}}{2
\left(1-\frac{d-1}2\da_1^2\right)}\,, \label{EMDLakRel1}\\
    -2\La_1\ell^2&=&(d-1)\left(d-\frac{d-1}{2}\da_1^2\right)\,. \label{EMDLaRel2}
\eea

\paragraph{Thermodynamics.}

Using the generalized reduction of the holographic conserved charges, one may compute various thermodynamic quantities, such as the temperature, entropy and mass of the black hole:
\bea
     T_{(d+1)} &=&  \frac{r_+^{1-\frac{d-1}2\d_1^2}}{4\pi\ell^2}\left[\left(d-\frac{d-1}2\d_1^2\right)+\frac{(d-2)\l_{(d-1)}\ell^{2}}{\left(1-\frac{d-1}{2}\d_1^2\right)r_+^{2-(d-1)\d^2}}-\frac{q^2\ell^{2}}{r^{2(d-1)}}\right],\\
    Q_{(d+1)}&=& \frac{V_{(d-1)}q}{16\pi G_N^{(d-1)}}\sqrt{2(d-1)\left(d-2+\frac{d-1}{2}\d_1^2\right)}\\
    \mu_{(d+1)}&=&\sqrt{\frac{2(d-1)}{(d-2+\frac{d-1}{2}\d_1^2)}}q r_+^{2-d-\frac{d-1}2\d_1^2}\,,\\
    S_{(d+1)}&=&\frac{V_{(d-1)}}{4G_N^{(d-1)}}r_+^{(d-1)},\\
    M_{(d+1)}&=&\frac{V_{(d-1)}(d-1)}{16\pi G_N^{(d-1)}}m.
\eea
One may check that the first law holds:
\be
    \ud M_{(d+1)}=T_{(d+1)}\ud S_{(d+1)}+\mu_{(d+1)}\ud Q_{(d+1)}\quad \Leftrightarrow \quad  \ud G_{(d+1)}=-S_{(d+1)}\ud T_{(d+1)}-Q_{(d+1)}\ud \m_{(d+1)}
\ee
and then examine global and local thermodynamical equilibrium in the grand-canonical ensemble by computing the Gibbs potential, the heat capacity and the electric permittivity\footnote{Both are straightforward to compute, but the expressions are cumbersome.}:
\bea
    G_{(d+1)}&=&\frac{V_{(d-1)}\left(1-\frac{d-1}2\d_1^2\right)r_+^{d-\frac{d-1}2\d_1^2}}{16\pi G_N^{(d-1)}}\left[-\ell^2+\frac{(d-2)\l_{(d-1)}r_+^{(d-1)\d_1^2-2}}{\left(1-\frac{d-1}2\d_1^2\right)\left(d-2+\frac{d-1}2\d_1^2\right)}-\right.\nn\\
    &&\qquad\qquad\qquad\qquad\qquad\qquad\qquad\left.-\frac{q^2}{r_+^{(d-1)(2-(d-1)\d_1^2)}}\right]\\
    C_\mu^{(d+1)}&=&\left.T\frac{\ud S}{\ud T}\right|_\m\,,\qquad \e_T^{(d+1)}=\left.\frac{\ud Q}{\ud \mu}\right|_T\,.
\eea
One finds out that when $\d_1^2\leq\d_c^2$, the thermodynamics is identical to that of a charged AdS black hole, \cite{Chamblin:1999tk,Chamblin:1999hg}.

\paragraph{Diagonal oxidation for $\gamma^2\leq\da^2_c$}

In this case, the lower-dimensional gauge field \eqref{GaugeEMDBH2Liouville} originates from a higher-dimensional Maxwell field strength in the action, as described above. Thus, from the result of section \ref{BHOxidation}, we can expect to recover the Reissner-Nordstr\"om solution in $(2\sig+1)$ dimensions, using the Ansatz \eqref{KKDiagonal}. This is indeed what happens and the solution \eqref{EMDBH2Liouville} uplifts to:
\bea
    \ud s^2_{(2\sigma+1)}&=&-\frac{f(\rho)}{\ell_{(2\s+1)}^{2}\r}\ud \tau^2 + \frac{\ell_{(2\s+1)}^{2}\ud \rho^2}{4\r^2f(\rho)}+\rho^{-1}\left(\ud X^2_{(d-1)}+\ud X^2_{(2\sig-d)}\right),\label{CAdSBHKK}\\
    f(\rho)&=&1+\ell_{(2\s+1)}^{2}\left( \lambda_{(2\sig-1)}\r-m\rho^{\sig}+q^2\rho^{2\sig-1}\right)\,,\\
    A&=&-\sqrt{\frac{2(2\sig-1)}{2\sig-2}}q\rho^{\sig-1}\ud \tau\,.\label{GaugeRNAdS}
\eea
To obtain this, we have used the same change of coordinates \eqref{AdSCoordChangeNeutral} as in section \ref{BHOxidation}, as well as rescaled the charge parameter $q\to(2\sig-1)q/(d-1)$.

The thermodynamic quantities become:
\bea
    T_{(2\s+1)}&=&\frac{\r_+^{-\frac12}}{4\pi}\left[\frac{2\s}{\ell_{(2\s+1)}^2}+(2\s-2)\l_{(2\s-1)}\r_+-(2\s-2)q^2\r_+^{(2\s-1)}\right], \\
    S_{(2\s+1)}&=&\frac{V_{(2\s-1)}}{4G_N^{(2\s+1)}}\r_+^{-\frac12(2\s-1)}\,, \\
    Q_{(2\s+1)}&=&\frac{V_{(2\s-1)}}{16\pi G_N^{(2\s+1)}}q\sqrt{2(2\s-1)(2\s-2)}\,,\\
    \mu_{(2\s+1)}&=&\sqrt{2\frac{2\s-1}{2\s-2}}q\r_+^{\s-1}\,,\\
    M_{(2\s+1)}&=& \frac{V_{(2\s-1)}}{16\pi G_N^{(2\s+1)}}(2\s-1)m\,,\\
    G_{(2\s+1)}&=& -\frac{V_{(2\s-1)}\r_+^{-\s}}{16\pi G_N^{(2\s+1)}}\left[\ell^2_{(2\s+1)}-\l_{(2\s-1)}\r_++q^2\r_+^{(2\s-1)}\right],
\eea
again, coinciding with results from \cite{Chamblin:1999hg,Chamblin:1999tk}.

When $\da_c^2 < \delta_1^2< \da_{max}^2$,  one can oxidize the solution
using (\ref{KKDiagonal2}) and  $\gamma=\d_2=\frac{2}{(d-1)\d_1}$
(which in particular means $\gamma^2\leq\da^2_c$).
This leads to the same solution (\ref{GaugeRNAdS}).

\subsubsection{Solution with a single exponential potential}

Let us now consider the case of the potential with a single exponential:
$\La_2=0$ and $\ga=\d_2=\d_c^2/\da_1$ in \eqref{EMD2LiouvilleAction}.
The field equations can be integrated to the following solution
(see \cite{Charmousis:2009xr} for its four-dimensional version):
\bsea
    \ud s^2_{(d+1)} &=& - e^{\ga\phi}\frac{V(p)}{p^2}\ud t^2+ \frac{e^{\da_1\phi}V(p)^{-1}\ud p^2}{(-\Lambda_1)\left(\d_{max}^2-\da_1^2\right)}
+  e^{\da_1\phi}\left(p-p_-\right)^{2\frac{\left(\d^2-\da_1^2\right)}{\left(\d_{max}^2-\da_1^2\right)}}\ud R^2_{(d-1)}\,, \slabel{Metric1} \\
    V(p) &=& (p-p_+)(p-p_-)\,, \slabel{Pot1} \\
    e^{\phi}&=& e^{\phi_0}p^{\frac{2\da_1}{\left[(d-2)\da_1^2
+\d_c^2\right]}}\left(p-p_-\right)^{\frac{2(d-1)\da_1\left(\da_1^2-\d_c^2\right)}{\left[(d-2)\da_1^2+\d_c^2\right]\left[\d_{max}^2-\da_1^2\right]}}\,, \slabel{Phi1}\\
    \mathcal A &=& \sqrt{\frac{2(d-1)\da_1^2p_-}{\left[(d-2)\da_1^2+\d_c^2\right]p_+}}\left(1-\frac{p_+}{p}\right)\ud t\,, \slabel{A1}\\
    \ga\da_1&=&\d_c^2\,,\qquad \d^2_c=\frac2{d-1}\,,\qquad \d_{max}^2=\frac{2d}{d-1}\,.\nn
    \label{SolGaDa1}
\esea

$p=0$ and $p_-$ are both curvature singularities, while $p_+$ is an event horizon. Setting the charge to zero, the neutral solution \eqref{EDBH2Liouville} is recovered with $\l_{(d-1)}\sim\La_2=0$.

The thermodynamics of this solution in $d+1=4$ were studied in \cite{Charmousis:2010zz}. These results generalize to higher dimensions straightforwardly. For the black hole solutions, that is when
\be
    \Lambda_1<0 \qquad \&\& \qquad \d_1^2<\d_{max}^2
    \label{BHCondition}
\ee
 holds, three ranges should be distinguished:
\begin{enumerate}
    \item Lower range: $\d_1^2\leq\d_c^2$. The solution behaves as the charged AdS planar black hole, there is a single locally stable branch, both in the canonical and grand-canonical ensembles.
    \item Middle range: $\d_c^2<\d_1^2<\frac{d^2-3}{(2d-3)(d-1)}+\frac{\sqrt{d+3}}{2d-3}$. The solution behaves as the charged (asymptotically flat) Reissner-Nordstr\"om black hole, there are two branches, small and large black holes, only the latter of which are locally stable.
    \item Upper range: $\frac{d^2-3}{(2d-3)(d-1)}+\frac{\sqrt{d+3}}{2d-3}\leq\d_1^2<\d_{max}^2$. The solution behaves as the (asymptotically flat) Schwarzschild black hole and is always locally unstable.
\end{enumerate}

\paragraph{Diagonal curved oxidation $\da^2_c<\d_1^2<\da^2_{max}$.}

In this range of values of $\d_1$, we expect that the uplift of \eqref{SolGaDa1} should give a charged, asymptotically flat version of the black brane \eqref{AdSBlackpBrane}. Indeed, using the diagonal Ansatz \eqref{KKDiagonal2} together with \eqref{BpBDaSigRelation} as in  \eqref{EMDdiagred2}, $\La_1$ plays the role of the curvature of the internal space instead of that of a higher-dimensional cosmological constant: this brings us to $(2\sig+1)$-dimensional Einstein-Maxwell theory, \emph{without cosmological constant}. After the following change of coordinates and identifications:
\be
    p=\rho^{2\sig-d-1}\,, \quad p_\pm=\rho_\pm^{2\sig-d-1}\,,
\ee
as well as
\be
    t = \sqrt{\lambda_{(2\sig-d)}} \tau\,, \quad -2\Lambda=(2\sig-d)(2\sig-d-1)\lambda_{(2\sig-d)}
\ee
the $(2\sig+1)$-dimensional solution is:
\bea
    \ud s^2_{(2\sigma+1)}&=&-f(\rho)\ud\tau^2+\left[1-\left(\frac{\rho_-}{\rho}\right)^{2\sig-d-1}\right]^{\frac{2(d-1)}{(2\sig-2)(2\sig-d-1)}}\left[\frac{\ud\rho^2}{f(\rho)}+\rho^2\ud K^2_{(2\sig-d)}\right]+ \nn\\
        &&\qquad+\left[1-\left(\frac{\rho_-}{\rho}\right)^{2\sig-d-1}\right]^{\frac{-2}{2\sig-2}}\ud R^2_{(d-1)}\,,\\
    f(\rho)&=& \lambda_{(2\sig-d)}\left[1 - \left(\frac{\rho_+}{\rho}\right)^{2\sig-d-1}\right]\left[1-\left(\frac{\rho_-}{\rho}\right)^{2\sig-d-1}\right],\\
     A&=&-\sqrt{\frac{2(2\sig-1)}{(2\sig-2)\lambda_{(2\sig-d)}}}\left(\frac{\rho_-\rho_+}{\rho^2}\right)^{2\sig-d-1}\ud\tau\,.
\eea
This solution can be better interpreted by going to
\be
    r^{2\sig-d-1}=\rho^{2\sig-d-1}-\rho_-^{2\sig-d-1}\,,
\ee
\be
     \rho_-^{2\sig-d-1}=r_0^{2\sig-d-1}\frac{\sinh^2 \omega}{\lambda_{(2\sig-d)}}\,,\quad \rho_+^{2\sig-d-1}=r_0^{2\sig-d-1}\frac{\cosh^2\omega}{\lambda_{(2\sig-d)}}\,,
\ee
we find
\bea
    \ud s_{(2\sig+1)}^2 &=& -K(r)^{-2}f(r)\ud\tau^2+K(r)^{\frac{2}{2\sig-2}}\left[\frac{\ud r^2}{f(r)}+r^2\ud X^2_{(2\sig-d)}+\ud R^2_{(d-1)}\right]
    \label{BlackqBrane2}\,,\\
    f(r)&=&\lambda_{(2\sig-d)}-\left(\frac{r_0}{r}\right)^{2\sig-d-1}\,,\qquad K(r)=1+\frac{\sinh^2\omega}{\lambda_{(2\sig-d)}}\left(\frac{r_0}{r}\right)^{2\sig-d-1}\,,\\
    A&=&-\sqrt{\frac{2(2\sig-1)\lambda_{(2\sig-d)}}{(2\sig-2)}}\left(1-K(r)^{-1}\right)\coth\omega\,\ud \tau\,.
\eea
 This is a $(d-1)$-brane supporting a point-like electric charge,
\cite{Gibbons:1994vm}. It can be obtained from the $(d-1)$-brane with
a $q$-charge (corresponding to a $(q+1)$-form potential in the
theory), where only $q\leq d -1$ directions of the worldvolume of the
brane support the charge, \cite{Caldarelli:2010xz}. Taking $q=0$
recovers the solution \eqref{BlackqBrane2}. It can also be obtained by
uplifting the asymptotically flat dilatonic black holes of
\cite{Gibbons:1987ps}.This points out an interesting relation between
asymptotically flat black holes and black branes with an exponential
potential, since they are mapped to each another by Kaluza-Klein
oxidation/reduction, depending on whether one reduces on the
worldvolume of the brane or on the compact space $\mathbf
X_{(2\sig-d)}$.

\subsection{Non-diagonal reduction to Einstein-Maxwell-Dilaton theories}

In this section, we consider adding a Maxwell gauge field to the action \eqref{ED2LiouvilleAction}, as in \eqref{EMD2LiouvilleAction}, this time generated in the reduction by turning on a Kaluza-Klein vector.

The metric Ansatz has an off-diagonal component
along one of the reduced directions. This reduction is consistent
(when only one scalar field is kept) only when the reduction is along  an
$\mathbf S^1$, see Appendix \ref{Appendix:NonDiagonalS1KK}:
\be \label{KKAnsatzNonDiag}
\ud s_{(d+2)}^2 = e^{-\da_1\phi}\ud s^2_{(d+1)} +e^{\frac\phi\da_1\left(\d_c^2-\da_1^2\right)}\left(\ud y+\mathcal A\right)^2,\qquad \mathcal A=\mathcal A_{M}\ud x^M\,.
\ee
As the reduction is only over a single dimension, one has to set
\be
    2\sig=d+1 \quad \Rightarrow \quad R_{(1)}=0
\ee
since the $\mathbf S^1$ has zero curvature. Then, the action \eqref{EMD2LiouvilleAction} is recovered with
\be
 \Lambda_2 = 0\,,\qquad \da_1=\sqrt{\frac{2}{d(d-1)}}<\d_c\,, \qquad \g=\d_2=\frac{\d_c^2}{\d_1}=\sqrt{\frac{2d}{d-1}}>\d_c\,. \label{EMDnondiagred1}
\ee
Alternatively, one may reduce with
\be \label{KKAnsatzNonDiag2}
\ud s_{(d+2)}^2 = e^{-\frac{\d_c^2}{\d_1}\phi}\ud s^2_{(d+1)} +e^{\frac\phi\d_1\left(\d_1^2-\da_c^2\right)}\left(\ud y+\mathcal A\right)^2,\qquad \mathcal A=\mathcal A_{M}\ud x^M\,,
\ee
and recover \eqref{EMD2LiouvilleAction} with
\be
 \Lambda_1 = 0\,,\qquad \da_1=\sqrt{\frac{2d}{(d-1)}}>\d_c\,, \qquad \g=\d_1=\sqrt{\frac{2d}{d-1}}>\d_c\,. \label{EMDnondiagred2}
\ee
Both reduction schemes yield $\g>\d_c$, and thus are complementary to those presented in section \ref{section:EMDdiagonal}. The price to pay is that there is a single exponential in the potential, and that the parameters $\d_1$, $\g$ have a fixed value.
We will discuss a more general reduction that allows for
generic $\sig$ in section \ref{section:GeneralKKreduction}.

We shall now consider the uplifts of the previous two charged solutions \eqref{EMDBH2Liouville} (restricted to a single exponential potential) and \eqref{Metric1}.

\paragraph{Non-diagonal oxidation to $(d+2)$ dimensions: $\d_1^2\geq\d_c^2$.}

Let us take the single-exponential restriction $\La_1=0$ of the solution \eqref{EMDBH2Liouville} and use the Ansatz \eqref{KKAnsatzNonDiag2} as well as \eqref{EMDnondiagred2}.
In that case, we expect to recover the Schwarzschild-AdS black brane \eqref{AdSBlackpBrane} carrying a wave, since now $\Lambda_2$ is identified with the higher-dimensional cosmological constant.
Let us call $r_\pm$ the roots of the black-hole potential $V(r)$ \eqref{PotEMDBH2Liouville}, which we can rewrite
\bea
     V(r)&=&-\lambda_{(d-1)}\frac{(d-2)}{2(d-1)^2}r^{2d}\left(1-\left(\frac{r_+}{r}\right)^{2d-2}\right)\left(1-\left(\frac{r_-}{r}\right)^{2d-2}\right),\\
    A &=& \sqrt{\frac{(-\lambda_{(d-1)})(d-2)}{2(d-1)^2}}\left(\frac{r_-}{r_+}\right)^{d-1}\left(1-\left(\frac{r_+}{r}\right)^{2d-2}\right)\ud t
\eea
where one has to keep in mind that, since $\gamma^2>\d_c^2$, $\lambda_{(d-1)}<0$ and hence there is an overall minus sign in $V(r)$.

After some manipulations,the higher-dimensional metric is:
\bea
    \ud s^2_{(d+2)} &=& -r^{2d-2}\left(1-\left(\frac{r_-}{r}\right)^{2d-2}\right)\left[\sqrt{\frac{(-\lambda_{(d-1)})(d-2)m}{2(d-1)^2r_+^{2d-2}}}\ud t-\sqrt{\frac{r_-^{2d-2}}{m}}\ud y\right]^2  +\nonumber \\
    &&+\frac{2(d-1)^2r^{-2}\ud r^2}{(-\lambda_{(d-1)})(d-2)\left(1-\left(\frac{r_+}{r}\right)^{2d-2}\right)\left(1-\left(\frac{r_-}{r}\right)^{2d-2}\right)}+\nonumber\\
    &&+\frac{r_+^{2d-2}}{m}r^{2d-2}\left(1-\left(\frac{r_-}{r}\right)^{2d-2}\right)\ud y^2 + \ud X_{(d-1)}^2\,,
\eea
where we have set $m=r_+^{2d-2}-r_-^{2d-2}$ and $\mathbf X_{(d-1)}$ is a negative curvature space, $\lambda_{(d-1)}<0$.

We can bring this last expression to a more standard form by changing to the coordinate
\be
    \rho^{-1}=r^{2d-2}-r_-^{2d-2}\,, \quad r_+^{2d-2}=m\cosh^2\omega\,,\quad r_-^{2d-2}=m\sinh^2\omega\,,
\ee
and replacing
\be
    d(d-2)\lambda_{(d-1)}=2\Lambda_2=2\Lambda\,.
    \label{HypCurvatureGaDa1}
\ee
Then:
\bea
    \ud s^2_{(d+2)} &=& -\r^{(-1)}\left(1-m\r\right)\left[\frac{\ud t}{\cosh\omega}-\sinh\omega\,\ud y\right]^2+\frac{\cosh^2\omega}{\r}\ud y^2   +\nonumber \\
    &&+\frac{d\ud \rho^2}{(-4\Lambda)\r^2\left(1-m\r\right)}+ \ud X_{(d-1)}^2\,.
\eea
This is just the boosted three-dimensional B(H)TZ black hole, \cite{Banados:1992gq,Banados:1992wn}, times a hyperbolic plane $\mathbf X_{(d-1)}$, whose curvature is fixed by \eqref{HypCurvatureGaDa1}.

\paragraph{Non-diagonal oxidation to $d+2$ dimensions: $\d_1^2\leq\d_c^2$.\label{Section:GaDa1NonDiagUplift}}

Let us uplift the solution \eqref{SolGaDa1} with the Ansatz \eqref{KKAnsatzNonDiag} and use \eqref{EMDnondiagred1}:
\bea
    \ud s^2_{(d+2)} &=& \frac{d\ud p^2}{(-2\Lambda)(d+1)V(p)}+(p-p^-)^{\frac2{d+1}}\ud R^2_{(d-1)}+\nn\\
    &&\qquad+p(p-p_-)^{-\frac{d+1}{d-1}}\left[\left(\ud y+\sqrt{\frac{p_-}{p_+}}\frac{p-p_+}p\ud t\right)^2-\frac{V(p)}{p^2}\ud t^2\right].
\eea
Changing coordinates,
\be
    p-p_-=r^{d+1}\,,\qquad p_+=m\cosh^2\omega\,,\quad p_-=m\sinh^2\omega\,,
\ee
leads to a more more standard form of the metric:
\bea
    \ud s^2_{(d+2)} &=& -\frac{r^2f(r)}{K(r)}\ud t^2+\frac{\ell^2_{(d+2)}\ud r^2}{f(r)}+r^{2}\ud R^2_{(d-1)}+r^2K(r)\left(\ud y+\tanh\omega\frac{f(r)}{K(r)}\ud t\right)^2,\\
    f(r)&=&r^2-\frac{m}{r^{d-1}}\,,\qquad K(r)=1+\frac{m\sinh^2\omega}{r^{d+1}}\,.
    \label{AdSwave}
\eea
A further change of the radial coordinate, $r^2=1/\r$, allows to recover the form of the metric \eqref{bbraneconfdt2} used in section \ref{Gen_br}.
After some manipulation of the $\ud y^2$ terms and rescaling $\bar y=\cosh\omega\, y$, the uplifted metric can be rewritten as:
\bea
    \ud s^2_{(d+2)} &=& -\frac{f(\r)}\r\left(\frac{\ud t}{\cosh\omega}-\tanh\omega\,\ud\bar y\right)^2+\frac{\ell^2_{(d+2)}\ud \r^2}{4\r^2f(\r)}+\frac1\r\left(\ud R^2_{(d-1)}+\ud \bar y^2\right),\\
    f(\rho)&=&1-m\r^{\frac12(d+1)}\,,
    \label{BoostedAdS}
\eea
which corresponds to Schwarzschild-AdS$_{d+2}$ carrying a wave. The cylindrical black string in four dimensions and its related stationary version have been studied in \cite{Lemos:1994xp,Lemos:1995cm}. The generalisation of the stationary cylindrical black hole to $d+1$ dimensions and to $[d/2]$\footnote{$[]$ means the integer part} arbitrary rotation parameters was presented in \cite{Awad:2002cz}. Making the change of coordinates:
\be
     t=\ch\omega~\bar t+\sh\omega~\bar y\,,
\ee
shows that the metric \eqref{BoostedAdS} is locally isometric the static black brane. This is only a local isometry because we are mixing one periodic coordinate ($y$) with the time coordinate. It is reflected in the fact that the first Betti number of this spacetime is not zero, and all closed curves are not in the same equivalence class, \cite{Stachel:1981fg} (the ones wrapped around the cylinder cannot be shrunk to a point).

If one unwraps the extra coordinate and takes its universal covering, then this stationary spacetime becomes globally isometric to the static AdS black brane, by boosting it along the worldvolume direction $y$. Now, a boost would usually mean the following coordinate transformation:
\be
    t = \ch\omega~ \bar t +\sh\omega~\bar y\,,\qquad  y = \sh\omega~\bar t +\ch\omega~\bar y\,.
\ee
One can show, reversing the previous steps, that this change of coordinates in the AdS black brane metric  gives back, after reduction along the boost direction, the solution \eqref{SolGaDa1} but where the gauge field  has been shifted so that it has zero chemical potential at spatial infinity.

Finally, boosted black branes in the context of the AdS/CFT correspondence were investigated in \cite{Cvetic:1998jf}. The thermodynamics for both this spacetime and its lower-dimensional reduction can be recovered from the formul\ae\ in section \ref{Gen_br}, by setting $2\s=d+1$ and subsequently turning off the extra scalar $\z$.

\subsection{Generalized non-diagonal reduction along a torus \label{section:GeneralKKreduction}}


The non-diagonal reduction discussed in the previous subsection was
restricted to the reduction over a single $\mathbf S^1$. We would
like to generalize the reduction to a torus reduction and then
consider a continuation over its dimension.
The generic, non-diagonal reduction of Einstein theory with a $4$-form field strength was performed in \cite{Lu:1995yn} for $D=11$
supergravity. It is straightforward to generalize the formul\ae\ to
$2\sig+1$ dimensions with a cosmological constant and we present
these results in this section. As in the two previous subsections,
straight latin capital fields refer to
higher-dimensional gauge fields, while calligraphic letters are reserved for
lower-dimensional gauge fields stemming from the reduction.

Our starting point is the AdS-Maxwell action in (\ref{EMLambdaAction})
and we would like to make a general non-diagonal torus reduction over
$\mathbf T^{(2 \sig-d)}$. The lower-dimensional fields are the metric,
$(2\sig{-}d)$ scalar fields $\vec{\phi}$ parametrizing the size of the
torus, 
gauge fields, $\mathcal A_{(1)}^a =\mathcal A_{(1)M}^a \ud x^M$ and
$A_{(1)} =A_{(1)M} \ud x^M$,
originating from the metric and higher-dimensional gauge field and
axions $\mathcal A_{(0)b}^a$
($a<b\leq2\sig-d$) and $A_{(0)a}$ ($a\leq 2\s-d$). The reduction ansatz is\footnote{The massive modes of the KK tower transform as doublets of the isometry group, while the massless modes transform as singlets. Since the isometry group is Abelian, the two representations do not mix. The massive modes are then not sourced by the massless modes, and can be safely truncated, \cite{PopeLectureNotes}.}
\bea
    \ud s^2_{(2\sig+1)} &=& e^{-\vec{\da}\cdot\vec{\phi}}\ud s^2_{(d+1)} + \sum_{a=d+1}^{2\sig}e^{-\vec{\gamma_a}\cdot\vec{\phi}}\left(h^a\right)^2\,,\label{GenKKAnsatz}\\
    h^a &=&\ud y^a + \mathcal A_{(1)}^a +
\sum_{b=a+1}^{2 \s} \mathcal A_{(0)b}^a\ud y^b\,,\\
     A_{(1)}^{(2\s+1)}&=&A^{(d+1)}_{(1)M}\ud x^M+A^{(d+1)}_{(0)a}\ud y^a\,,\qquad F_{(2)}^{(2\s+1)}=\tilde F_{(2)}^{(d+1)}+\tilde F_{(1)a}^{(d+1)}h^a\,,\label{GenGaugeKKAnsatz}
\eea
where we have explictly made the distinction between higher- and lower-dimensional gauge fields. We define the tilded fields just below.
This leads to a reduced theory governed by the action
\bea
    S_{(d+1)}&=&\frac{1}{16\pi G_N^{(d+1)}}\int\ud^{d+1}x\,\sqrt{-g}\left[R-\half\left(\partial\vec{\phi}\right)^2-2\La e^{-\vec{\da}\cdot\vec{\phi}}-\frac14\sum_{a}e^{-\overrightarrow{\b_a}\cdot\vec{\phi}}\left(\tilde{\mathcal F}_{(2)}^a\right)^2\right.\nn\\
         &&\left.-\frac12\sum_{a<b}e^{-\vec{\b_{ab}}\cdot\vec{\phi}}\left(\tilde{\mathcal F}_{(1)}^{ab}\right)^2-\frac14e^{\vec{\da}\cdot\vec{\phi}}\left(\tilde F_{(2)}\right)^2-\frac12\sum_{a}^{}e^{-\vec{\a_{a}}\cdot\vec{\phi}}\left(\tilde{F}_{(1)a}\right)^2\right], \label{KKGenAction}\\
    \tilde{\mathcal F}_{(2)}^a &=& \mathcal F_{(2)}^a   - \gamma^b_{\phantom{2}c}\mathcal F_{(1)b}^a\wedge\mathcal A_{(1)}^c\,,\qquad  \mathcal F_{(2)}^a=\ud \mathcal A_{(1)}^a\\
    \tilde{\mathcal F}_{(1)b}^{a} &=&\gamma^c_{\phantom{2}b}\mathcal F_{(1)c}^{a}\,,\qquad\qquad\qquad\qquad \mathcal F_{(1)b}^{a}=\ud \mathcal A_{(0)b}^{a},\\
    \tilde F_{(2)}&=& F_{(2)} - \tilde F_{(1)a}\wedge\mathcal A_{(1)}^a\,,\qquad\quad  F_{(2)}=\ud A_{(1)}\\
    \tilde F_{(1)a}&=& \gamma^b_{\phantom{2}a} F_{(1)b}\,,\qquad\qquad\qquad\quad\quad  F_{(1)}=\ud A_{(0)}\,,\\
  \gamma^a_{\phantom{2}b}&=&\left[(1+\mathcal A_{(0)})^{-1}\right]^a_{\phantom{2}b} = \da^a_{\phantom{2}b} - \mathcal A_{(0)b}^a+ \mathcal A_{(0)b}^c \mathcal A_{(0)c}^a+\ldots
\eea
where the tilded field strengths include extra transgression terms\footnote{It is the tilded field strengths which compare directly with their higher-dimensional counterparts, as can be seen from \eqref{GenGaugeKKAnsatz}.} compared to their untilded, usual definition as the exterior derivative of the gauge potential. The axionic metric $\gamma^a_{\phantom{2}b}$ has a finite number of terms since a particular axion $\mathcal A_{(0)b}^a$ is only defined for $a<b$ and
$\vec{\delta}, \vec{\gamma}_a, \vec{\b_a},  \vec{\a_a},
\vec{\b_{ab}}$ are given by:
\bea
    \vec{\da}&=& \left(\da_{d+1},\ldots,\da_{2\sig}\right),\qquad   \da_a=\sqrt{2/((2\sig-a)(2\sig-a-1))}\, \\
     \vec{f_a}&=&\left(\underset{a-1}{\underbrace{0,\ldots,0,}}(2\sig-a)\da_{d+a},\da_{d+1+a},\ldots,\da_{2\sig}\right),\\
    \vec{\gamma_a}&=&\vec{\da}-\vec{f_a} =  \left(\da_{d+1},\ldots,-(2\sig-a-1)\da_{d+a},\underset{2\sig-d-a}{\underbrace{0,\ldots,0}}\right),\label{coef} \\
    \vec{\a_a}&=&\vec{f_a}-\vec{\da}\,,\qquad  \vec{\b_a}=-\vec{f_a}\,,\qquad \vec{\b_{ab}}=-\vec{f_a}+\vec{f_b}\,.
\eea

Let us now work out how we may generate solutions to the equations of motion stemming from \eqref{KKGenAction}. We may start from the charged planar AdS black hole and boost it along $2\sig-d$ directions of the horizon, \cite{Awad:2002cz}, with $\omega_a$ the boost parameters:
\bea
    \ud s^2_{(2\sig+1)}&=&-\frac{f(\rho)}\r\left(\xi\ud\tau-\sum_{a=1}^{2\sig-d}\omega_a\ud y^a\right)^2+\frac{\rho^{-1}}{\ell_{(2\sig+1)}^4}\sum_{a=1}^{2\sig-d}\left(\omega_a\ud\tau-\xi\ell^2_{(2\sig+1)}\ud y^a\right)^2-\nn\\
                && -\frac{\rho^{-1}}{\ell_{(2\sig+1)}^2}\sum_{a<b}\left(\omega_a\ud y^b-\omega_b\ud y^a\right)^2 + \rho^{-1}\ud R_{(d-1)}^2+\frac{\ud\rho^2}{4\r^2f(\rho)}\,,\\
    f(\rho)&=&\ell_{(2\s+1)}^{-2}-m\rho^{\sig}+q^2\rho^{2\sig-1}\,,\\
    A&=&-\sqrt{\frac{2(2\sig-1)}{2\sig-2}}q\rho^{\sig-1}\left(\xi\ud\tau-\sum_{a=1}^{2\sig-d}\omega_a\ud y^a\right),\\
    \xi&=&1+\sum_{a=1}^{2\sig-d}\frac{\omega_a^2}{\ell_{(2\s+1)}^2}\,.
\eea
It is now a matter of calculation to show that all the terms in the $(\tau,y^a)$ sector can be rearranged as
\bea
    \ud s^2_{(2\sig+1)}&=&-\frac{f(\rho)}{\r K_{2\sig-d}}\ud\tau^2 + \rho^{-1}\ud R_{(d-1)}^2+\frac{\ud\rho^2}{4\r^2f(\rho)}+\sum_{a}\frac{K_{a}}{\r K_{a-1}}\left(h^a\right)^2\,,\\
    h^a&=&\ud y^a-\frac{\omega_a}{\sum_{b\leq a}\omega_b^2}\left(1-K_a^{-1}\right)\left(\xi\ud\tau-\sum_{b>a}\omega_b\ud y^b\right),\\
    K_{a}(\rho)&=&1+\sum_{b\leq a}\omega_b^2\left(m\rho^{\sig}-q^2\rho^{2\sig-1}\right)\,,
\eea
where by convention $\omega_0=0$ and thus $K_0=1$. This is precisely the form we need to match the Kaluza-Klein reduction Ansatz \eqref{GenKKAnsatz}. Thus, one can show that the $(d+1)$-dimensional theory \eqref{KKGenAction} admits the following solution
\bea
    \ud s^2_{(d+1)}&=&-\frac{\rho^{-\frac{2\sig-1}{d-1}}f(\rho)}{\left(K_{2\sig-d}\right)^{\frac{d-2}{d-1}}}\ud\tau^2+\rho^{-\frac{2\sig-d}{d-1}}\frac{\left(K_{2\sig-d}\right)^{\frac1{d-1}}}{4\r^2f(\rho)}\ud\rho^2+ \rho^{-\frac{2\sig-1}{d-1}}\ud R_{(d-1)}^2\,,\\
     e^{\phi_a}&=&\rho^{-\frac12(2\sig-1)\da_i}\left(K_a\right)^{\frac1{(2\sig-a-1)\da_a}}\left(K_{a-1}\right)^{\frac{-1}{(2\sig-a)\da_a}},\\
    \mathcal A^a_{(1)}&=&-\frac{\omega_a}{\sum_{b\leq a}\omega_b^2}\left(1-\left(K_a\right)^{-1}\right)\xi\ud\tau\,,\\
    \mathcal A^a_{(0)b} &=&\frac{\omega_a\omega_b}{\sum_{b\leq a}\omega_b^2}\left(1-\left(K_a\right)^{-1}\right)\,,\\
    A_{(1)}&=&-\sqrt{\frac{2(2\sig-1)}{2\sig-2}}q\rho^{\s-1}\xi\ud\tau\,,\\
    A_{(0)a}&=&\sqrt{\frac{2(2\sig-1)}{2\sig-2}}q\rho^{\sig-1}\omega_a\,.
\eea
If one wishes to delete any kind of reference to the higher-dimensional theory from which this solution originates, it suffices to replace $2\sig=d+\mathcal N$, where $\mathcal N$ is now simply the number of calligraphic gauge fields. Let us stress at this point that the procedure by which we obtained the above solution is not quite trivial, since there was no guarantee that we could reach the Kaluza-Klein form from the boosted black brane.

We may not analytically continue $\mathcal N$ to arbitrary real values yet, since we cannot analytically continue the number of gauge fields! One may remedy this by
using a mixed diagonal/non-diagonal reduction: first, reduce diagonally along $\mathcal N-\mathcal M$ dimensions, then non-diagonally along $\mathcal M$ dimensions. One can then continue analytically the number of diagonal dimensions, \emph{i.e.} $\mathcal N-\mathcal M$. In practice, setting
\be
    \forall\,a,b \leq \mathcal N-\mathcal M \quad  \mathcal A^a_{(1)}=\mathcal A^a_{(0)b} =0  \quad \Rightarrow \quad h^a=\ud y^a,
\ee
as well as
\be
    \frac{d+\mathcal M-1}{2}\da^2=\frac{\mathcal N-\mathcal M}{d+\mathcal N-1}\Leftrightarrow \mathcal N=\frac{\mathcal M+(d-1)(d+\mathcal M-1)\frac{\da^2}2}{1-(d+\mathcal M-1)\frac{\da^2}2}\,,
\ee
and
\be
    e^{\frac{\Phi}\da\left(\frac2{d+\mathcal M-1}-\da^2\right)}= e^{-\vec{\gamma_a}\cdot\vec{\phi}}\quad \forall\,a\leq \mathcal N-\mathcal M
\ee
gives
\be
    \sum_{a=1}^{\mathcal N-\mathcal M}\phi_a^2 = \Phi^2\,.
\ee
The reduction Ansatz now is
\bea
    \ud s^2_{(2\sig+1)} &=& e^{-\vec{\da}\cdot\vec{\phi}-\da\Phi}\ud s^2_{(d+1)} +e^{-\da\Phi}\ud R_{(\mathcal N-\mathcal M)}^2+ \sum_{a}^{\mathcal M}e^{-\vec{\gamma_a}\cdot\vec{\phi}}\left(h^a\right)^2\,,\\
    \da_a&=&\sqrt{\frac2{(\mathcal M+d-a)(\mathcal M+d-a-1)}} \quad a=1\ldots\mathcal  M\,,\\
     \vec{f_a}&=&\left(\underset{a-1}{\underbrace{0,\ldots,0,}}(\mathcal M+d-a)\da_{d+a},\da_{d+1+a},\ldots,\da_{\mathcal M+d}\right),\\
    \vec{\gamma_a}&=&\vec{\da}-\vec{f_a} =  \left(\da_{d+1},\ldots,-(\mathcal M+d-a-1)\da_{d+a},\underset{\mathcal M-a}{\underbrace{0,\ldots,0}}\right),
\eea
where the diagonal reduction runs over $\mathcal N-\mathcal M$ dimensions and the non-diagonal one over $\mathcal M$ and the various (arrowed) vectors are $\mathcal M$-dimensional. The $(d+1)$-dimensional action is
\bea
    S_{(d+1)}&=&\frac{1}{16\pi G_N^{(d+1)}}\int\ud^{d+1}x\,\sqrt{-g}\left[R-\half\partial\Phi^2- \half\left(\partial\vec{\phi}\right)^2-2\La e^{-\vec{\da}\cdot\vec{\phi}-\da\Phi}\right.\nn\\
        &&-\frac14\sum_{a}^{\mathcal M}e^{-\vec{\b_a}\cdot\vec{\phi}}\left(\tilde{\mathcal F}_{(2)}^a\right)^2-\frac12\sum_{a<b}^{\mathcal M}e^{-\vec{\b_{ab}}\cdot\vec{\phi}}\left(\tilde{\mathcal F}_{(1)}^{ab}\right)^2 \nn\\
        &&\left.-\frac14e^{\vec{\da}\cdot\vec{\phi}+\da\Phi}\left(\tilde F_{(2)}\right)^2-\frac12\sum_{a}^{\mathcal M}e^{\da\Phi-\vec{\a_{a}}\cdot\vec{\phi}}\left(\tilde{F}_{(1)a}\right)^2\right],
\eea
which has a solution
\bea
    \ud s^2_{(d+1)}&=&-\frac{\rho^{-\frac{2\sig-1}{d-1}}f(\rho)}{\left(K_{2\sig-d}\right)^{\frac{d-2}{d-1}}}\ud\tau^2+\rho^{-\frac{2\sig-d}{d-1}}\frac{\left(K_{2\sig-d}\right)^{\frac1{d-1}}}{4\r^2f(\rho)}\ud\rho^2+ \rho^{-\frac{2\sig-1}{d-1}}\ud R_{(d-1)}^2\,,\\
    e^{\Phi}&=&\rho^{-\frac12\frac{(d+\mathcal M-1)\da}{1-\frac{d+\mathcal M-1}2\da^2}}\,,\\
    2\sig&=&\frac{\mathcal M+d-(d+\mathcal M-1)\frac{\da^2}2}{1-(d+\mathcal M-1)\frac{\da^2}2}\,,\\
     e^{\phi_a}&=&\rho^{-\frac12(2\sig-1)\da_a}\left(K_a\right)^{\frac1{(2\sig-a-1)\da_a}}\left(K_{a-1}\right)^{\frac{-1}{(2\sig-a)\da_a}},\quad a=1\ldots \mathcal M\\
    \mathcal A^a_{(1)}&=&-\frac{\omega_a}{\sum_{b\leq a}\omega_b^2}\left(1-\left(K_a\right)^{-1}\right)\xi\ud\tau,\quad a=1\ldots \mathcal M\,,\\
    \mathcal A^a_{(0)b} &=&\frac{\omega_a\omega_b}{\sum_{b\leq a}\omega_b^2}\left(1-\left(K_a\right)^{-1}\right),\quad a,b=1\ldots \mathcal M\,,\\
    A_{(1)}&=&-\sqrt{\frac{2(2\sig-1)}{2\sig-2}}q\rho^{\sig-1}\xi\ud\tau\,,\\
    A_{(0)aa}&=&\sqrt{\frac{2(2\sig-1)}{2\sig-2}}q\rho^{\sig-1}\omega_a,\quad a=1\ldots\mathcal M\,.
\eea
Since $\da$ is now taken to be any real number, so is $2\sig$, and we have thus generated a \emph{family} of solutions depending on the real parameter $\da$. In the following, we shall restrict the simplest case $\mathcal M=1$, $q=0$, with only one gauge field generated from the metric, two Kaluza-Klein scalars and no axion or higher-dimensional Maxwell field.

\section{Holography from generalized  dimensional reduction} \label{Gen_Hol}

We will now consider the simplest non-trivial case and work it out completely.
This is the $\mathcal M=1$, $q=0$ case discussed in the previous section
(one gauge field, two Kaluza-Klein scalars and no axion or
higher-dimensional Maxwell field). We start
by deriving the generalized Kaluza-Klein reduction map in the next
subsection. Although the map has already been given in the previous section,
the discussion in this section serves as an illustration of the steps
involved in its derivation. Furthermore, we will also be able to connect
more directly with the discussion in \cite{Kanitscheider:2009as}.
Then we will move on and use these results to derive the holographic dictionary,
followed by the derivation of the universal holographic hydrodynamics.

\subsection{Generalized dimensional reduction} \label{subsec:genred}

We start from Einstein gravity with negative cosmological constant
in $(2 \s + 1)$ dimensions and consider a reduction that involves
a Kaluza-Klein gauge field,
\be \label{AdS_up}
S_{(2 \s+1)} = L_{AdS} \int \ud^{2 \s+1} x \sqrt{-g_{(2 \s+1)}}
\left[R + 2 \s (2 \s-1)\right].
\ee
where  $L_{AdS} =  \ell_{(2 \sig+1)}^{2\s-1}/(16 \pi G_{2\s+1})$,  $\ell_{(2 \sig+1)}$ is the
AdS radius and we used an appropriate Weyl rescaling to move $\ell_{(2 \sig+1)}$
as an overall constant in the action.

We use the following reduction ansatz for the theory on the torus $\mathbf T^{(2 \s-d)}$
\be
\label{redans}
\ud s_{(2 \sigma +1)}^2 = \ud s^2_{(d+1)}(\r,z) + e^{2 \phi_1(\r,z)} \left(\ud y  - A_M \ud x^M\right)^2+
e^{\frac{2\phi_2(\r,z)}{(2\s-d-1)}} \ud y^a \ud y^a,
\ee
where $a=1,\ldots ,(2\s-d-1)$. The coordinates $(y,y^a)$ are periodically
identified with period $2 \pi R$ and $x^M=(\rho,z^i)$ with $M=0,\ldots, d$.
This is a consistent truncation, since the resulting lower-dimensional field equations are equivalent
to the higher-dimensional field equations. The resulting lower-dimensional theory is governed by the action
\bea
S_{(d+1)} &=& 
L \int \ud ^{d+1} x \sqrt{-g_{(d+1)}} e^{\f_1+\f_2} \left[R
+ 2 \partial \f_1 \partial \f_2 + \frac{2 \s -d-2}{2 \s-d-1} (\partial \f_2)^2
\right.\nonumber \\
&& \qquad \qquad \left.-\frac{1}{4} e^{2 \phi_1} F_{MN} F^{MN} + 2 \s (2 \s-1)
\right]. \label{Ricactred}
\eea
where  $L = L_{AdS} (2 \pi R)^{2\s-d}$.

One can derive this action as follows. First reduce on the $(2 \s - d  -1)$-dimensional
torus to obtain
\be \label{red-first}
S_{(d+2)} = L_{AdS} (2 \pi R)^{2 \s -d -1}
 \int \ud ^{d+2} x \sqrt{-g_{(d+2)}} e^{\f_2} \left[R_{(d+2)} + \frac{2 \s -d-2}{2 \s-d-1} (\partial \f_2)^2
+ 2 \s (2 \s-1) \right].
\ee
To obtain this result we use the fact that
\be
   R_{(2\s+1)} = R_{(d+2)} - 2 \na^2 \phi_2 - \frac{2\s- d}{2\s-d - 1} (\pa \phi_2)^2.
\ee
Now we reduce on the $y$ direction including a Kaluza-Klein gauge field. Note
that
\be
\ud s^2_{(d+2)} = \ud s^2_{(d+1)} + e^{2 \phi_1} \left(\ud y - A_{M} \ud x^{M}\right)^2,
\ee
and thus
\be
R_{(d+2)} = R_{(d+1)} - 2 \na^2 \phi_1 - 2 (\pa \phi_1)^2  - \frac{1}{4} e^{2 \phi_1} F_{MN} F^{MN}.
\ee
Substituting into \eqref{red-first} leads to \eqref{Ricactred}.
%
Setting $F_{MN} = 0$, and rescaling
\be
\phi_1 = \frac{\phi}{(2 \s -d)}; \qquad
\phi_2 = \frac{\phi (2 \s  -d  -1)}{( 2 \s -d)},
\ee
with $\phi = (\phi_1 + \phi_2)$ results in the action  for non-conformal branes derived in \cite{Boonstra:1998mp,Kanitscheider:2008kd},
\be \label{lowdaction}
S = L
\int \ud ^{d+1} x \sqrt{-g_{d+1}} e^{\f} \left(R
+ \frac{2 \s -d-1}{2 \s-d} (\partial \f)^2 + 2 \s (2 \s  -1 ) \right ).
\ee


It is natural to rewrite the action \eqref{Ricactred} in terms of the scalar
\be
\psi = \phi_1 + \phi_2,
\ee
since the determinant of the metric over the torus is expressed in terms of $\psi$ as $\sqrt{g_{T^{2\s -d}}} = e^{\psi}$. We also
use the combination
\be
\zeta = (2 \s - d  -1 ) \f_1 - \f_2,
\ee
in terms of which the reduction of the metric is
\be \label{red_DF}
\ud s^2_{(2\s+1)} =\ud s^2_{(d+1)} + e^{2 \frac{(\psi + \zeta)}{(2 \sigma -d)}} (\ud y - A_{M}\ud x^{M})^2 + e^{\frac{2 \psi}{(2 \s-d)} -
\frac{2 \zeta}{(2\s-d)(2 \s -d -1)}} \ud y^a \ud y_a,
\ee
and the action becomes
\bea
S_{(d+1)} &=&  L
\int \ud^{d+1} x \sqrt{-g_{(d+1)}} e^{\psi} \left[ R
- \frac{1}{(2 \s -d)(2\s - d-1)} (\partial \zeta)^2
\right.\nonumber \\
&& \qquad \qquad \left.
+ \frac{2 \s -d-1}{2 \s-d} (\partial \psi)^2
-\frac{1}{4} e^{\frac{2 (\zeta + \psi)}{(2 \s-d)}} F_{MN} F^{MN} + 2 \s (2 \s-1)
\right]. \label{Ricactred2}
\eea
Note that the equation of motion for $\zeta$ is
\be
\nabla [e^{\psi} \pa \zeta] = \frac{1}{4} (2 \s  -d -1)e^{\psi}  e^{\frac{2 (\zeta + \psi)}{(2 \s-d)}} F_{MN} F^{MN}.
\ee
This implies that is always consistent to set $\zeta = 0$ when $F_{MN} =0$: the action with both $\zeta$ and $F$ set
to zero is precisely that given above, with the identification $\psi = \phi$.

The equation of motion for $\psi$ is
\bea
\nabla [e^{\psi} \pa \psi] &=& \frac{2\s -d}{2(2\s -d -1)}e^{\psi}\left[ R -\frac{1}{(2\s-d)(2\s-d-1)}(\pa \zeta)^{2}+\frac{2\s-d-1}{2\s-d}(\pa \psi)^{2} \right . \nono \\
&& \left . -\frac{1}{4}\frac{(2\s -d+2)}{(2\s-d)}e^{\frac{2(\zeta+\psi)}{(2\s-d)}}F_{MN}F^{MN} + 2\s(2\s-1)  \right],
\eea
and the gravitational field equation is
\bea
&& R_{MN}  - \frac{1}{2}g_{MN}R -\frac{1}{(2\s -d)(2\s -d -1)}\left( -\frac{1}{2}g_{MN}(\pa\zeta)^{2}+\pa_{M}\zeta\pa_{N}\zeta\right) \nono \\
&+& \frac{1}{2\s -d}\left( \frac{1}{2}(2\s-d+1)g_{MN}(\pa\psi)^{2} - \pa_{M}\psi\pa_{N}\psi \right) \\
&-& \frac{1}{4}e^{\frac{2(\zeta + \psi)}{2\s-d}}\left( -\frac{1}{2}g_{MN}F_{PQ}F^{PQ} + 2F_{M}^{\, Q}F_{NQ} \right) -\s(2\s-1)g_{MN} \nono \\
&-& \nabla_{N}\nabla_{M}\psi +g_{MN}\Box \psi = 0. \nono
\eea
The gauge field equation is
\be
\nabla_{M}  \left ( e^{\frac{2 (\zeta + \psi)}{(2 \s-d)}} F^{MN} \right ) = 0.
\ee
From the equations of motion, one can notice that there are certain special values of $\sigma$. In the case of $2 \sigma = (d+1)$, the reduction is along a circle
and there is no additional scalar field $\zeta$. The case of $2\sigma = d$ corresponds to the conformal case where there is no reduction at all, and one retains only the metric. One should note that it is also clearly consistent to set the gauge field to zero whilst retaining both scalars $(\zeta,\psi)$. These field equations are, as mentioned above, completely equivalent to the higher-dimensional Einstein equations so the reduction is consistent.

We may conformally rescale the action to bring it into Einstein frame, using the following rescaling of the metric,
\be
g_{MN} = e^{-2\psi/(d-1)} \bar{g}_{MN}
\ee
 to get:
\bea
S_{(d+1)} &=&  L
\int \ud^{d+1} x \sqrt{-\bar{g}_{(d+1)}} \left[ \bar{R}
- \frac{1}{(2 \s -d)(2\s - d-1)} (\partial \zeta)^2
+ \frac{1-2 \s}{(2 \s-d)(d-1)} (\partial \psi)^2
\right.\nonumber \\
&& \qquad \qquad \left.-\frac{1}{4} e^{\frac{2 (\zeta + \psi)}{(2 \s-d)} + \frac{2\psi}{d-1}} F_{MN} F^{MN} + e^{-\frac{2\psi}{d-1}} 2 \s (2 \s-1)
\right].
\eea
Note that the potential is clearly independent of the scalar $\zeta$.
In order to obtain canonically normalized scalar kinetic terms we rescale
the scalars as
\be
\psi = \sqrt{\frac{(2\s-d)(d-1)}{2(2\s-1)}} \bar{\psi},\qquad
\zeta = \sqrt{\frac{(2\s-d)(2\s-d-1)}{2}} \bar{\zeta},
\ee
to yield the action in the Einstein frame
\bea
{S}_{(d+1)} &{=}& {L \int\ud^{d+1} x \sqrt {-\bar{g}_{(d+1)}}}
{ \left[ \bar{R} - \frac{1}{2} (\pa \bar{\psi})^2
- \frac{1}{2} (\pa \bar{\zeta})^2
+ 2\s(2\s-1) e^{-\bar{\psi} \sqrt{\frac{2(2\s-d)}{(d-1)(2\s - 1)}}}
\right.}\nono \\
&& { \left. - \frac{1}{4} e^{\sqrt{\frac{2(2\s-1)}{(d-1)(2\s-d)}} \bar{\psi} + \sqrt{\frac{2(2\s-d-1)}{2\s-d}} \bar{\zeta} } F_{MN} F^{MN} \right]}.
\label{red_action}
\eea
Note that this rescaling implicitly assumes that $2 \sigma > (d+1)$: the scalar $\zeta$ has a negative kinetic term whenever $2 \sigma < (d+1)$ and therefore
cannot be canonically normalized. For such values of the parameter $\sigma$, one would not expect that the scalar $\zeta$ is part of a physical compactification, as we discuss next\footnote{It is interesting to note that a similar action was recently discussed in \cite{Chemissany:2011gr}, in the context of $p$-branes with curved worldvolumes. However, the scalar potentials in this case are different, and one cannot interpret the action given here in terms of branes with curved
worldvolumes.}.

\subsubsection{Brane interpretation}

In this section we discuss whether the $(d+1)$-dimensional action
(\ref{red_action}) can be interpreted in terms
of consistent truncations of sphere reductions of decoupled D$p$-brane,
M-brane and string solutions. Let us begin by reviewing the case with the
metric and one scalar, $\psi$, discussed in
\cite{Boonstra:1998mp,Kanitscheider:2008kd}.
 If one truncates the action \eqref{red_action} to just these fields, and sets
\be
2 \sigma = d + \frac{ (p-3)^2}{(5-p)},
\ee
with $d = p + 1$ and $p \neq 5$, then the action arises from the reduction of the corresponding decoupled $p$-brane background over a sphere.
The scalar field $\psi$ is then dual to the running coupling of the worldvolume theory, whilst the metric is dual to the field theory energy momentum tensor. The general
parametrization encompasses the conformal cases of the D3-branes, M2-branes and M5-branes, with the latter M-branes obtained by setting $2\sigma = d$.
It also includes the cases of D0-branes, D1-branes, D2-branes, D4-branes
and fundamental strings, with the latter corresponding to $p=1$ in the
formula above but excludes five-branes and D$p$-branes with $p \ge 6$.

For the non-conformal cases of the D1-branes, D4-branes and fundamental strings,
\be
(2 \sigma - d) = 1,
\ee
which implies that the action \eqref{Ricactred2} can always be interpreted as an $S^1$ reduction of a conformal theory. In this case the scalar $\zeta$ is not present, as the only reduction is the standard KK reduction over a circle. The gauge field in these cases is just the Kaluza-Klein gauge field of the reduction, corresponding to the
conserved current in the reduced field theory.

For the case of D2-branes, notice that
\be
(2 \sigma - d)  = 1/3 < 1,
\ee
which implies that the kinetic term (in Einstein frame) in \eqref{Ricactred2}
for the new scalar $\zeta$ is negative (or if we work with \eqref{red_action}
the coefficient of $F^2$ becomes complex and the action is not real).
Decoupled D2-branes reduced on an $S^6$ are believed to admit a consistent truncation \cite{Boonstra:1998mp} which is the $ISO(7)$ gauged supergravity theory
\cite{Hull:1984yy,Hull:1984vg}. The corresponding operators to these gauged supergravity fields would be the operators in the same supermultiplet as the stress energy tensor. The gauged supergravity theory contains however no scalars with negative kinetic terms, and therefore $\zeta$ cannot be interpreted as one of the scalars of the gauged supergravity theory, nor indeed would it seem to have a sensible interpretation in terms of the dual (supersymmetric) gauge theory.

The final standard case is that of the D0-branes, for which
\be
(2 \sigma - d) = \frac{9}{5}.
\ee
In this case the scalar $\zeta$ has a positive kinetic term, and there is no a priori obstruction to it being interpreted as one of the scalars arising in an $\mathbf S^8$ compactification of the type IIA theory in ten dimensions. At the same time, there is also no guarantee that the scalar $\zeta$ and the gauge field can be identified with fields in the $\mathbf S^8$ reduction.

\bigskip

There is a second natural way to interpret the $(d+1)$-dimensional actions in terms of decoupled branes and strings: given the action corresponding to a non-conformal $p$-brane in $d' + 1 \equiv (p+2)$ dimensions, one can always reduce this action on a circle to obtain an action with an additional scalar and gauge field in one less dimension. In such cases, the relation between the index $\sigma$ and $p$ would be
\be  \label{sigma_value}
2 \sigma = p + 2  + \frac{(p-3)^2}{(5-p)},
\ee
with $p \neq 5$, and the dual theory is the KK reduction of the
non-conformal $p$-brane theory.

\subsection{Holographic dictionary}

We now want to use the generalized dimensional reduction in order to set up
a holographic dictionary for this theory. In general, in order to set up such
a dictionary one needs to understand the asymptotic structure of the field
equations, which is a hard problem. We have just established however
that all solutions of the theory (\ref{red_action}) descend from solutions
of (\ref{AdS_up}) and the most general asymptotic solution of the latter
is known:
\bea \label{AdS_exp}
\ud s_{(2 \s +1)}^2 &=& \frac{\ud \rho^2}{4 \rho^2} + \frac{1}{\rho} g_{\mu \nu} \ud x^{\mu} \ud x^{\nu}; \\
g_{\mu \nu} &=& g_{(0) \mu \nu} + \rho g_{(2) \mu \nu} + \cdots + \rho^{\sigma} \left ( g_{(2 \sigma) \mu \nu} + h_{(2 \sigma) \mu \nu} \log \rho \right ) + \cdots,
\eea
where $g_{(0) \mu \nu}$ is the source,
only the trace and divergence of $g_{(2 \sigma) \mu \nu}$ are determined
locally in terms of the source and all other coefficients
are completely determined. The logarithmic terms
$h_{(2 \sigma)}$ are present only when $\sigma$ is integral.
It follows that it suffices to consider the class of asymptotic solutions that
is also of the form (\ref{red_DF}) required for the reduction in order
to obtain the general asymptotic solution of \eqref{Ricactred2}.

The $(d+1)$-dimensional metric is expanded in the usual Fefferman-Graham
form, as above, whilst
the scalar fields can be expanded as
\bea \label{AdS_exp_scalars}
e^{\frac{2 \psi}{(2 \sigma -d)}} &=& \frac{1}{\rho} e^{\frac{2 \kappa}{(2 \sigma -d)}}; \qquad \kappa = \kappa_{(0)} + \rho \kappa_{(2)}
+ \cdots + \rho^{\s} \kappa_{(2\s)}; \nn \\
\zeta &=& \zeta_{(0)} + \rho \zeta_{(2)} + \cdots
+ \rho^{\sigma} \zeta_{(2 \s)},
\eea
and the gauge field as
\be \label{AdS_exp_gauge}
{ A_i (\r,z) = A_{i(0)}(z) + \r A_{i(2)}(z) + ... + \r^\s A_{i(2\s)}(z) + ... }
\ee
There are log terms present when $\s$ is an integer, but these are once
again suppressed, as we are primarily interested in the cases where
$\s$ is non-integral. Here $\kappa_{(0)}, \zeta_{(0)}$ and $A_{(0)}$ are the
sources to dual scalar operators, $\cO_\psi$, $\cO_\zeta$ and the
conserved current $J^i$. The subleading coefficients are locally related to
the sources, up to the order where the vev of the dual operator appears.
The precise form can be worked out from the known local relation between
the subleading coefficients in (\ref{AdS_exp}) and $g_{(0)}$ (see
appendix A of \cite{deHaro:2000xn}), but we will not need these relations here.

Having obtained the asymptotic solution, one would then next like to
compute the local boundary counterterms that would render finite the
on-shell action. Happily, this can be easily done using the
generalized dimensional reduction \cite{Kanitscheider:2009as}.
Given $\sigma$ we choose any half-integer $\tilde{\s}> \s $ and determine
the $[\s]+1$ most singular AdS$_{(2\tilde{\s}+1)}$-counterterms as a
function of $\tilde{\s}$, where $[\s]$ denotes the largest integer less
than or equal to $\s$ (when $\s$ is an integer one of these counterterms
is logarithmic). Reducing these AdS$_{(2\tilde{\s}+1)}$-counterterms
and replacing $\tilde{\s}$ by $\s$ yields the counterterms appropriate
for \eqref{Ricactred2}.

As an example let us consider the counterterm action for $1 < \s < 2$,
for which
we only need two counterterms. The two most singular counterterms in AdS$_{2\tilde{\s}+1}$ defined
on a regulating hypersurface are given by
(see appendix B of \cite{deHaro:2000xn})\footnote{Note that convention for the curvature tensor used in \cite{deHaro:2000xn} has the opposite sign.}
\be
\label{Sct2sigma}
   S^{ct}_{(2\tilde\s)} = L_{AdS} \int_{\r=\e} \ud^{2\tilde{\s}}x \sqrt{-\g_{2\tilde{\s}}} \left[2(2\tilde{\s}-1) + \frac{1}{2\tilde{\s}-2} \hat{R}[\g_{2\tilde{\s}}]\right],
\ee
where $\g_{2\tilde{\s} ij}$ is the induced metric on the $(2\tilde{\s})$-dimensional hypersurface and $\hat{R}[\g_{2\tilde{\s}}]$ the corresponding curvature.
The counterterm action to \eqref{Ricactred2} for $1< \s < 2$ is then given by reducing \eqref{Sct2sigma} to $d$ dimensions and replacing $\tilde{\s}$ with $\s$,
\bea
   S^{ct}_{(d)} &=&  L \int_{\r=\e} \ud^dx \sqrt {-\g_d}\,\, e^{\psi} \left[2 (2\s-1) + \frac{1}{2\s-2} \left(\hat{R}_d
+ \frac{2\s -d -1}{2\s-d} (\pa \psi)^2  \right . \right. \\
&& \qquad \left . \left . -  \frac{1}{(2\s -d -1)(2\s-d)} (\pa \zeta)^2
- \frac{1}{4} e^{\frac{2(\zeta + \psi)}{2\s-d}} F_{ij} F^{ij}  \right)\right]. \nn
\eea
This covers early results for $d=3$, \cite{Cai:1999xg}. When $\s > 2$ one needs to include additional gravitational counterterms.

Next let us turn to holographic one point functions. These can be
computed by functionally differentiating the renormalized on-shell
action, $S_{ren}$,  but again the dimensional reduction offers a shortcut:
we simply need to reduce the formula for the 1-point function.
The latter reads \cite{deHaro:2000xn},
\be
\label{conf_dict}
   \<T_{\m\n}\>_{2\s} = \frac{2}{\sqrt{-g_{(0),2\s}}} \frac{ \d S_{ren}}{\d g_{(0)}^{\m\n}} = 2\s L_{AdS}g_{(2\s) \m\n}  + \ldots,
\ee
where the ellipses denote terms that locally depend
on $g_{(0)\m\n}$. These terms are present when $g_{(0)\m\n}$ is curved and there
is a conformal anomaly, \emph{i.e.} when $\s$ is an integer. They do not play an
important role in the discussion here and so they will be suppressed.

Reduction of the expectation value of the higher-dimensional stress
energy tensor gives the expectation values of the operators in the
$d$-dimensional field theory. Let us begin by writing the former in
terms of components longitudinal and transverse to the reduction
torus. When we do so, we should also take into account the additional
prefactor $(2\pi R)^{2\s-d}$ of the lower-dimensional action in
\eqref{Ricactred} which results from the integration over the torus
and for the change in the determinant of the metric in the definition
of the vev, $\sqrt{g_{(0),d}} = e^{-\k\sub{0}} \sqrt{g_{(0),2\s}}$.
To accommodate these factors, we define
\be
\<t_{\mu \nu}\>_d \equiv  e^{\k\sub{0}} (2\pi R)^{2\s-d} \<T_{\mu \nu}\>_{2\s}\,.
\ee
Then one obtains
\bea
\<t_{ij}\>_d  &=& 2\s L \left[ e^{\k\sub{0}} g_{(2\s)ij} + 2
e^{\frac{ (2 \s - d +2)\k\sub{0} + 2 \zeta\sub{0}}{2 \s-d}}
\left(A_{(i(0)} A_{j)(2\s)}  
+ \frac{A_{i(0)}A_{j(0)}}{2\s-d} \left(\k\sub{2\s} + \zeta\sub{2 \s}\right)\right )  \right],
\nono  \\
\< t_{iy} \>_d   &{ = }& {- 2\s L
e^{\frac{ (2 \s - d +2)\k\sub{0} + 2 \zeta\sub{0}}{2 \s-d}}}
\left(A_{i(2\s)} + \frac{2}{2\s-d} \left(\k\sub{2\s} + \zeta\sub{2\s} \right) A_{i(0)}
\right),  \label{vev} \\
\< t_{yy} \>_d   &=& \frac{4 \s L}{(2 \s -d)}
e^{\frac{ (2 \s - d +2)\k\sub{0} + 2 \zeta\sub{0}}{2 \s-d}}
\left(\k\sub{2\s} + \zeta\sub{2\s}\right) + \ldots
\equiv - e^{\frac{2}{(2\s-d)} \left(\k\sub{0} +\zeta\sub{0}\right)}
\<\cO_1\>_d, \nono \\
\< t_{ab} \>_d   &=& \frac{4\s L}{2\s -d}
e^{\frac{1}{2 \s-d}
\left((2 \s - d +2)\k\sub{0} - \frac{2}{(2 \sigma -d -1)}\zeta\sub{0}\right)}
\left(\k\sub{2\s}  - \frac{1}{(2 \s - d -1)} \zeta\sub{2\s} \right)\d_{ab} + \ldots
\nono \\
   &\equiv & - e^{\frac{2}{2 \s-d}\left( \k\sub{0} - \frac{1}{(2 \sigma -d -1)}
\zeta\sub{0}\right)}
\<\cO_2\>_d
\d_{ab}, \nono
\eea
where the ellipses again contain curvatures of the boundary metric $g_{(0)ij}$ and
derivatives of $(\k_{(0)}, \zeta_{(0)})$. From these expressions we read off
\bea
\<\cO_1 \>_d&=& - \frac{4\s L}{2\s-d} e^{\k\sub{0}}
\left(\k\sub{2\s} + \zeta\sub{2\s}\right) + \ldots,  \\
\<\cO_2 \>_d &=& - \frac{4\s L}{2\s-d} e^{\k\sub{0}}
\left(\k\sub{2\s}  - \frac{1}{(2 \s - d -1)} \zeta\sub{2\s} \right) + \ldots \nn
\eea
The reduction gives, as expected, a symmetric tensor operator $t_{ij}$, a vector operator $t_{iy}$ and two scalar operators.
The normalizations of all the operators at this point is somewhat arbitrary with the stress energy tensor, current and naturally normalized
scalar operators of the dual $d$-dimensional field theory being formed from linear combinations of these operators. The combinations which form the
$d$-dimensional field theory operators can be obtained by varying the renormalized onshell actions with respect to the appropriate sources, for example
the stress energy tensor follows from varying the action with respect to the $d$-dimensional metric source. There is however a simple way to deduce the appropriate
combinations from the reduction of the higher-dimensional Ward identities. Anticipating how this reduction will work, let us introduce linear combinations of the scalar
operators such that
\bea
\<  \cO_{\psi} \>_d &=& \frac{1}{(2\sigma -d)} \left [ (2 \sigma - d -1) \<  \cO_{2} \>_d + \<\cO_{1}\>_{d} \right ]; \\
\< \cO_{\zeta} \>_d &=& \frac{1}{(2 \sigma -d)} \left [ \<  \cO_{1} \>_d - \<\cO_{2} \>_{d} \right ]. \nn
\eea
The notation follows from the fact that the field $\psi$ will act as a source for ${\cO_{\psi}}$ whilst the field $\zeta$ sources
${\cO_{\zeta}}$. It is useful to recall the
$\zeta = 0$ limit. In this case note that $\<\cO_1 \>_d = \<\cO_2 \>_d$, with the operator $\<\cO_\phi\>_d$
defined in \cite{Kanitscheider:2008kd} taking the expectation value
\be
\<\cO_\phi\>_d = \<\cO_1 \>_d = \<\cO_2 \>_d.
\ee

The conformal Ward identity $\<T^\m_\m\>_{2\s} = \cA_{2\s}$ in the $2\s$-dimensional theory can be reduced to
\bea
   && {\<t^i_i\>_{d} -2A_{(0)}^{i}\< t_{iy}\>_{d} - (2 \s - d -1) \< \cO_{2}\>_d - \left(1
+ e^{\frac{2 \left(\k_{(0)} + \zeta_{(0)}\right)}{2\s - d}} A_{(0) i} A_{(0)}^{i}\right)\< \cO_{1}\>_{d}} \nono \\
   && {= e^{\k\sub{0}} (2\pi R)^{2\s -d}\cA_{2\s} \equiv \cA_d\,.}
\eea
Furthermore, if we write
\bea
\label{redforms}
    &&
\< {J_{i} \>_{d} = \< t_{iy}\>_{d} + A_{(0)i} \<t_{yy}\>_{d}}\,,  \label{com} \\
    && {\< {T}_{ij} \>_{d} = \<t_{ij}\>_{d} + \left(A_{(0) i} \< J_{j} \> + A_{(0)j} \< J_{i} \>\right)
+A_{(0)i} A_{(0) j} e^{\frac{2 \left(\k_{(0)} + \zeta_{(0)}\right)}{2\s - d}}\<\cO_{1}\>_{d}}\,, \nono
\eea
so that
\bea
&& \< J_{i} \>_{d} = -2 \s L
e^{\frac{1}{2 \s-d}\left( (2 \s - d +2)\k\sub{0} + 2 \zeta\sub{0}\right)}
A_{(2\s)i} + \cdots; \\
&& \< {T}_{ij} \>_{d} = 2 \s L e^{\k\sub{0}} g_{(2\s)ij} + \cdots, \nn
\eea
{the dilatation Ward identity becomes simply
\be \label{dil_WI}
{\<{T}_{i}^{i}\>_{d} - (2 \s  -d ) \<\cO_{\psi}\>_{d} = \cA_{d}}.
\ee
Note in particular that the new scalar operator $\cO_{\zeta}$ does not contribute to the dilatation Ward identity.

Using these linear combinations of the operators, the conservation equation for the higher-dimensional stress energy tensor reduces to
\be\label{ConsEqT}
\na^i \< {T}_{ij} \>_{d} + \pa_j \k\sub{0} \< \cO_\psi \>_{d} + \pa_j \zeta\sub{0} \< \cO_\zeta \>_{d} - F_{(0)j}^{i}\< J_{i}\>_{d}= 0,
\ee
{and} the divergence equation for a current
\be \label{ConsEqJ}
{\na^i \< J_{i} \>_{d} = 0.}
\ee
Looking at the first divergence equation we can recognize it as the standard diffeomorphism Ward identity for a theory with stress energy tensor $T_{ij}$
in which the other operators are defined in terms of the generating functional $W$
\be
\< J^i \>_d= -\frac{1}{\sqrt{g_{(0)}}} \frac{\delta W}{\delta A_{(0)i}}; \qquad
\< \cO_{\psi} \>_d = - \frac{1}{\sqrt{g_{(0)}}} \frac{\delta W}{\delta \k\sub{0}}; \qquad 
\< \cO_{\zeta} \>_d = - \frac{1}{\sqrt{g_{(0)}}} \frac{\delta W}{\delta \zeta\sub{0}},
\ee
indicating that the non-normalizable modes of $(\psi,\zeta)$ do indeed source $ (\cO_{\psi}, \cO_{\zeta})$ respectively, as anticipated,
whilst $A_{(0)i}$ sources the conserved current $J^i$. One could directly verify these relations by varying the renormalized bulk onshell action.

\subsection{Black branes} \label{Gen_br}

In preparation of our discussion of hydrodynamics in the next subsection,
we will now discuss a realization of the setup in the previous section
using black branes. Recall that conformal hydrodynamics was
derived in \cite{Bhattacharyya:2008jc} by studying the long wavelength
fluctuation equations around the boosted black D3 brane geometry. The
universal hydrodynamics related to the non-conformal branes is
similarly related to long wavelength fluctuation equations around the
boosted black D$p$ brane geometry and can be most easily obtained by
starting from the (conformal) black brane solution in $(2 \s +1)$
dimensions and then carrying out the generalized dimensional reduction
\cite{Kanitscheider:2009as}. Here the starting point is a (conformal) black
brane solution in $(2 \s +1)$ dimensions:
\bea
\label{AdSbbrane}
   \ud s^2_{(2\s+1)} &=& \frac{\ud \r^2}{4\r^2 f(\rho)}
+\frac1\r\left[-f(\r)\ud t'^2 + \ud y'^2 + \ud z_r \ud z^r+ \ud y_a \ud y^a\right], \\
   f(\r) &=& 1 - m^{2\s} \r^\s\,, \nn
\eea
where $(y,y^a, z^r)$ run over all transverse coordinates ($a=d+1, \ldots, 2\s-1$). This metric is Einstein with negative curvature when $2 \s $ is an integer, and has an event horizon at $\r=m^{-2}$.  The Hawking temperature $T$ and Bekenstein-Hawking entropy density $s$ are given by
\be \label{entropyAdSbbrane}
T = \frac{ m \sigma}{2 \pi}, \qquad s = 4 \pi L_{AdS}  m^{2 \sigma-1}.
\ee

Performing a Lorentz transformation $t=\cosh\w~t'-\sinh\w~y'\,,\,y=\cosh\w ~y'-\sinh\w~t'$, the resulting metric can carry a wave: 
\bea
\label{bbraneconfdt2}
   \ud s^2_{(2\s+1)} &=& \frac{\ud \r^2}{4\r^2 f(\rho)}
-  \rho^{-1} K(\r)^{-1} f(\r)\ud t^2 + \frac{K (\r)}{\r} \left[\ud y - \left((K' (\r))^{-1}-1\right) \ud t\right]^2 \nn \\
&& +\ \rho^{-1} \ud z_r \ud z^r+\ \rho^{-1} \ud y_a \ud y^a, \\
   f(\r) &=& 1 - m^{2\s} \r^\s, \qquad K(\r) = \left(1 + Q \r^{\s}\right), \nn \\
\left(K'(\r)\right)^{-1} &=& \left(1 - \bar{Q} \r^{\s} K(\r)^{-1}\right), \nn
\eea
where 
\be
Q = m^{2 \s} \sinh^2 \omega; \qquad \bar{Q} = m^{2 \s} \sinh \w \cosh \omega.
\ee
Setting $\omega = 0$ removes the wave, whilst the extremal limit is recovered in the limit $m \rightarrow 0$
with $\omega \rightarrow \infty$ and $Q$ finite. 
When $(2 \sigma + 1)$ is integral, this solution arises
from a standard non-extremal intersection  of one of the conformal branes (D3, M2, M5) with a wave (see for example \cite{Cvetic:1996gq, Skenderis:1999bs}),
taking a decoupling limit (which focuses the geometry near the brane) and then reducing over the transverse sphere. The physical
interpretation of cases in which $(2 \sigma +1)$ is non-integral will be discussed in the next section.

With a view towards dimensional reduction, we consider now the coordinates
$(y,y^a)$ periodically identified with period $2 \pi R$ 
(as in subsection (\ref{subsec:genred})),
and 
rewrite the geometry
as
\bea
\ud s_{(2\s+1)}^2 &=& \frac{\ud\r^2}{4\r^2 f(\rho)} + \frac{1}{\rho} \left(\ud z^r \ud z_r - \ud t^2\right) + \frac{1}{\rho} \left(1 - K(\rho)^{-1} f(\rho)\right) \ud t^2 \\
&& \qquad+ \frac{1}{\rho}\ud y^a \ud y_a + \frac{K(\rho)}{\rho} \left[\ud y -\left((K' (\r))^{-1}-1\right) \ud t\right]^2. \nn
\eea
Now let us boost this geometry along the non-compact boundary dimensions with boost parameter $\hat u_i$ (where now $z^{i} = (t,z^r)$, {\it i.e.} the $y$ and $y^{a}$ directions are excluded). This
results in
\bea \label{boostbbranedep2}
\ud s_{(2\s+1)}^2 &=& \frac{\ud\r^2}{4\r^2 f(\rho)} + \frac{1}{\rho} \left(\ud z^{i} \ud z_{i}\right) + \frac{1}{\rho} \left(1 - K(\rho)^{-1} f(\rho)\right) \hat u_{i}\hat u_{j} \ud z^{i} \ud z^{j} \\
&& \qquad + \frac{1}{\rho}\ud y^a \ud y_a+ \frac{K(\rho)}{\rho} \left[\ud y- \left((K' (\r))^{-1}-1\right)\hat u_{i} \ud z^{i}\right]^2. \nn
\eea
Note that the fluid velocity $\hat u^i$ does \emph{not} square to $-1$ with the $2\s$-dimensional boundary metric, but $\eta_{ij}\hat u^i\hat u^j=-1$. In what follows, we shall also include an external, uniform gauge field $A_{(0)i}\ud z^i$, which can be obtained from \eqref{boostbbranedep2} by performing a coordinate transformation on $y$ as $\ud y\to\ud y+A_{(0)i}\ud z^i$. Once we allow the temperature, charge, fluid velocity and external gauge field to become position dependent the metric needs to be corrected at each order
to satisfy the field equations.

 We now reduce \eqref{boostbbranedep2}. The reduced metric is then
\be
\ud s_{(d+1)}^2 = \frac{\ud\r^2}{4\r^2 f(\rho)} + \frac{1}{\rho} \left(\ud z^{i} \ud z_{i}\right) + \frac{1}{\rho} \left(1 - K(\rho)^{-1} f(\rho)\right) \hat{u}_{i} \hat{u}_{j} \ud z^{i} \ud z^{j},
\ee
with the scalar fields being
\be
e^{\frac{2 \phi_2}{(2 \s - d - 1)}} = \frac{1}{\rho}, \qquad e^{2 \phi_1} = \frac{K(\rho)}{\rho},
\ee
and the gauge field is
\be \label{ReducedGaugeField}
A = \left[A_{(0)i}+ \left((K' (\r))^{-1}-1)\right) \hat u_{i}\right] \ud z^{i}.
\ee
Rewriting the scalar fields in terms of $(\psi,\zeta)$ we obtain
\be
e^{\psi} = \frac{1}{\rho^{\sigma - d/2}} K(\rho)^{1/2}, \qquad
e^{\zeta} = K(\rho)^{\frac{1}{2} (2 \s  -d -1)}.
\ee
It is useful to rewrite quantities using Fefferman-Graham coordinates
(see \cite{Kanitscheider:2009as} for the derivation of the coordinate
transformation).
The reduced metric is then
\bea \label{ReducedAdSbbrane}
\ud s_{(d+1)}^2 &=& \frac{\ud\tilde{\r}^2}{4\tilde{\r}^2} + \frac{1}{\tilde{\r}} \left( 1 + \frac{m^{2\s}\tilde{\r}^\s}{4} \right)^{\frac{2}{\s}} \ud z_i \ud z^i  \\
&& \ \ \ \ \ \ \ \ + \frac{1}{\tilde{\r}} \left( 1 + \frac{m^{2\s}\tilde{\r}^\s}{4} \right)^{\frac{2}{\s}}
 \left[ 1 - K(\r(\tilde{\r}))^{-1} f(\r(\tilde{\r})) \right] \hat{u}_i \hat{u}_j \ud z^i \ud z^j. \nn
\eea
with the scalar fields being
\be
e^{\frac{2 \phi_2}{(2 \s - d - 1)}} = \frac{1}{\tilde{\r}} \left( 1 + \frac{m^{2\s}\tilde{\r}^\s}{4} \right)^{\frac{2}{\s}} , \qquad
e^{2 \phi_1} = \frac{K(\r(\tilde{\r}))}{\tilde{\r}} \left( 1 + \frac{m^{2\s}\tilde{\r}^\s}{4} \right)^{\frac{2}{\s}} ,
\ee
and the gauge field is
\be \label{gaufie}
A = A_{(0)i}\ud z^i+ \left [ \left( K^{'}(\r(\tilde{\r})) \right)^{-1} - 1\right]\hat{u}_i \ud z^i.
\ee
Again rewriting the scalar fields in terms of $(\psi,\zeta)$ we obtain
\be
e^{\psi} = \frac{K(\r(\tilde{\r}))^{\half}}{\tilde{\r}^{\frac{2\s-d}{2}}} \left( 1 + \frac{m^{2\s} \tilde{\r}^\s}{4} \right)^{\frac{2\s-d}{\s}}, \qquad
e^{\zeta} = K(\r(\tilde{\r}))^{\frac{2\s-d-1}{2}}.
\ee
Now since
\be
e^{\frac{2\psi}{2\s-d}} = \frac{1}{\tilde{\r}} e^{\frac{2\k}{2\s-d}} \nn
\ee
we get that
\be
e^{\k} = K(\r(\tilde{\r}))^{\half} \left( 1 + \frac{m^{2\s}\tilde{\r}^\s}{4} \right)^{\frac{2\s-d}{\s}}.
\ee
We can expand the above results in $\tilde{\r}$ to get
\bea
\label{expansions}
\k_{(0)} = 0 &;& \k_{(2\s)} = \half Q + \frac{2\s-d}{\s} \frac{m^{2\s}}{4}, \\
\zeta_{(0)} = 0 &;& \zeta_{(2\s)} = \frac{2\s-d-1}{2} Q\,, \nn \\
&&  A_{i(2\s)} = \hat{u}_i \bar{Q}. \nn
\eea
The source for the gauge field is $A_{i(0)}$, i.e. the term in square brackets
in (\ref{gaufie}) goes to zero as $\tilde{\rho} \to 0$.

These allow us to extract the expectation values of the dual operators using \eqref{vev}. One finds,
\bea
\label{dualop}
\< {T}_{ij} \>_d &=& L m^{2\s} \eta_{ij} + 2\s L(Q + m^{2\s}) \hat{u}_i \hat{u}_j \\
                     &=& L m^{2\s} \left(\eta_{ij} + 2\s\cosh^2 \omega \hat{u}_i \hat{u}_j \right) ; \nn \\
\< J_{i} \>_d            &=& 2\s L \bar{Q} \hat{u}_i \nn \\
                     &=& 2\s L m^{2\s} \sinh \omega \cosh \omega \hat{u}_i ; \nn \\
\< {\cal O}_{1} \>_d &=& - m^{2\s} L - 2\s LQ \nn \\
                       &=& -L m^{2\s} \left( 1 + 2\s \sinh^2 \omega \right) \nn \\
\< {\cal O}_{2} \>_d &=& -L m^{2\s}; \nn
\eea
which one can verify indeed satisfies the dilatation Ward identity (\ref{dil_WI}). From these expressions we can also read off the thermodynamic quantities,
\be \label{thermo}
\hat \epsilon= L m^{2\s} (2 \s \cosh^2 \w -1), \qquad \hat q \equiv \bar{Q} =  2\s L m^{2\s} \sinh \omega \cosh \omega, \qquad \hat P=L m^{2\s},
\ee
where $\hat \epsilon$ is the energy density, $\hat q$ the charge density and $\hat P$ the pressure of the reduced spacetime \eqref{ReducedAdSbbrane}. One may solve the first two
equations to express $m$ and $\w$ in terms of $\hat \e$ and $\hat q$ and then use them in the last relation to obtain the equation of
state $\hat P=\hat P(\hat \e,\hat q)$,
\be \label{pressure}
\hat P(\hat \e,\hat q)=\frac{1}{2 \sigma-1} \left(\sqrt{\hat \e^2 (\s-1)^2 +(\hat \e^2- \hat q^2) (2 \s-1)} - \hat \e (\s-1)\right).
\ee
(Since $\hat P=L m^{2\s}$ this relation also expresses $m$ in terms of $\hat \e, \hat q$, while $\sinh 2 \w = \hat q/(\s \hat P)$ gives $\w$ in terms of $\hat \e$ and $\hat q$).
In the limit $\hat q \to 0$ we get the equation of state for the non-conformal branes.
Above extremality, $\hat \e > |\hat q|$, the expression under the square root is manifestly positive. In the
extremal limit $\hat \e \to |\hat q| $ the pressure vanishes, as expected.
The reduced temperature and entropy are
\be \label{entropyReduced}
\hat T = \frac{ m \sigma}{2 \pi \cosh \w}, \qquad \hat s = 4 \pi L \cosh \omega m^{2 \sigma-1}.
\ee

From the equation of state \eqref{pressure}, we may obtain the adiabatic 
speed of sound\footnote{See \cite{landau:1987:fm}, Chapter XV, equation (134.14) and (134.7).}
\be
	\hat c_s^2=\left.\frac{\partial \hat P}{\partial\hat \epsilon}\right|_{\hat s/\hat q}\,,
\ee
keeping fixed the ratio $\hat s/\hat q$. Using \eqref{entropyReduced} and \eqref{thermo}, this yields
\be
 	\ud \left(\frac{\hat s}{\hat q}\right)=0 \Rightarrow \ud\omega=-\tanh\omega\frac{\ud m}m
\ee
so that
\be \label{speedofsound}
	\hat c_s^2=\frac1{2(\s-1)\cosh^2\w+1}
\ee
which reduces to the result for the neutral black branes derived in \cite{Kanitscheider:2009as}.

Furthermore, from (\ref{ReducedGaugeField}) we obtain that the chemical 
potential is equal to
\be \label{chemical}
\hat \m = -\left(\left.\hat u^iA_i\right|_{\r=0}-\left.\hat u^iA_i\right|_{\r=m^{-2}}\right) = \tanh\w\,.
\ee
Regularity at the horizon requires that $\hat u^iA_i|_{\r=m^{-2}}=0$
which then fixes the external gauge field in terms of the chemical potential.
We will however relax this condition so that we can incorporate
a general external gauge field in the next subsection. We note, however, 
that all of
the main results (transport coefficients etc.) can equally be obtained
without turning on an additional external field beyond that required by the 
presence of the chemical potential. 
One may also verify that the thermodynamic identities,
\be
\hat P+\hat \e = \hat T \hat s + \hat q \hat \m, \qquad \ud \hat P = \hat s \ud \hat T + \hat q \ud \hat \m
\ee
hold.

It is interesting to observe that
the expectation values of the scalar operators, $(\< {\cal O}_{\psi} \>_d, \< {\cal O}_{\zeta} \>_d)$, can be expressed in terms of the energy density and pressure as
\bea \label{dualop1}
\< {\cal O}_{\psi} \>_d &=& \frac{1}{(2 \sigma -d)} \< {T}_{i}^i \>_d = \frac{1}{(2 \sigma -d)} \left [ (d-1) \hat P - \hat \ep \right ]; \\
\< {\cal O}_{\zeta} \>_d &=& \frac{1}{(2 \sigma -d)} \left [ (2 \sigma -1) \hat P - \hat \ep \right ]. \nn
\eea
Thus the expectation value of the scalar operator $\< {\cal O}_{\psi} \>_d$ characterizes the deviation of the equation of state from conformality (as one would expect)
whilst the expectation value of the second operator $\< {\cal O}_{\zeta} \>_d$ is zero in the uncharged case, in which case the equation of state indeed reduces to that of the non-conformal branes, $\hat P = \hat \ep/(2 \sigma -1)$.

\subsection{Universal Hydrodynamics}

We would now like to use the generalized dimensional reduction in order to
obtain the universal hydrodynamics corresponding to the charged
dilatonic solutions. Recall that
the hydrodynamic energy-momentum tensor for a conformal fluid at first-derivative order in $(2\s)$ dimensions on a curved manifold with metric $g\sub{0}_{\m\n}$ is
\bea
\label{hydroconfTmn}
  \< T_{\mu \nu} \>_{2\s}&=& \<T_{\mu \nu}^{\rm eq}\>_{2\s} + \<T_{\mu \nu}^{\rm diss}\>_{2\s} \\
  \< T_{\mu \nu}^{\rm eq}\>_{2\s}&=&P(g\sub{0}_{\mu \nu} + 2\s u_{\mu} u_{\nu}), \qquad
  \<T_{\mu \nu}^{\rm diss}\>_{2\s}= - 2\eta_{2\s} \s_{\mu \nu}, \nono \\
   \s_{\mu \nu} &=& P_{\mu}^{\kappa} P_{\nu}^{\lambda} \na_{(\kappa} u_{\lambda)}
   - \frac{1}{2\s-1} P_{\mu \nu} (\na \cdot u), \qquad P_{\mu \nu} = g\sub{0}_{\mu \nu} + u_{\mu} u_{\nu}, \nono
\eea
where $T$, $u_{\mu}$ and $\eta_{2\s}$ denote the temperature, velocity and shear viscosity respectively of the fluid and $\na_{\mu}$ is the covariant derivative corresponding to the metric $g\sub{0}_{\mu \nu}$. Note that we are working in Landau-Lifshitz frame,
\be
\label{LLcons}
u^{\mu}\<T^{\rm diss}_{\mu \nu}\>_{2\s}=0\,.
\ee
The evolution of the fluid is determined by the conservation of the energy-momentum tensor,
\be
\label{convcons}
   \na^{\mu} \<T_{\mu \nu}\>_{2\s} = 0\,.
\ee
For the AdS black brane,
\be \label{PressureAdSbbrane}
	P=L_{AdS} m^{2\s}\,, \qquad \eta_{2 \s}=\frac{s}{4\pi}=L_{AdS}m^{2\s-1}
\ee
by \eqref{entropyAdSbbrane}.

Let us first determine the reduced fluid velocity. The boundary metric can be read off the reduction Ansatz \eqref{red_DF}, using the expansions of the fields \eqref{AdS_exp}, \eqref{AdS_exp_scalars} and \eqref{AdS_exp_gauge}. For simplicity, we set $\k_{(0)}=\z_{(0)}=0$ as in the case of the AdS black brane \eqref{expansions}. Then,
\be \label{BoundaryMetric}
g_{(0)i j}=\h_{ij}+A_{(0) i}A_{(0) j}, \qquad
g_{(0)i y}= -A_{(0) i}, \qquad g_{(0)yy}=1\,,
\ee
and the inverse metric is given by
\be \label{BoundaryInverseMetric}
g_{(0)}^{ij} = \eta^{ij}, \qquad g_{(0)}^{i y} = A_{(0)}^i, \qquad g_{(0)}^{yy} =1+\eta^{ij}A_{(0) i}A_{(0) j}\,.
\ee
Note that the reduced boundary metric is simply the Minkowski metric $\eta_{ij}$\footnote{The hydrodynamics at first-order is independent of the curvature of the reduced boundary metric, so our results will still hold at first-order for a curved boundary in $d$ dimensions.}. One may then derive the reduced fluid velocity $\hat u^i$ by requiring that both
\be
u^{\mu} u_{\mu} =-1, \qquad u^{\mu} = g_{(0)}^{\mu \nu} u_{\mu},
\ee
and
\be
\hat u^{i} \hat u_{i} =-1, \qquad \hat u^{i} = \h^{ij} \hat u_{j},
\ee
It is convenient to choose by convention (and to make a link with the wave generating coordinate transformation of the previous subsection)
\be
	u_{y} = \sinh\w
\ee
so that, setting $ u_a=0$ along the remaining compact dimensions $y^a$,
\be \label{u_red}
 u_{i} = \cosh \w\hat u_i-\sinh\w A_{(0)i}\,,\quad u_{y} = \sinh\w\,, \quad u^i=\cosh\w\hat u^i\,,\quad u^y=\sinh\w+\cosh\w\hat u\cdot\partial A_{(0)}\,.
\ee

We may now turn to the equilibrium part. 
Inserting in $\<T_{\m\n}^{\rm eq}\>_{2\s}$ and using \eqref{redforms} we obtain:
\bea
\label{dualop2}
\< T^{\rm eq}_{ij} \>_d &=& \hat P \left[\eta_{ij} + 2\s\left(u_i+u_yA_{(0)i}\right)\left(u_j+u_yA_{(0)j}\right)\right], \nn \\
\< J_{i}^{\rm eq} \>_d  &=& 2\s\hat Pu_{y}\left(u_i+ u_{y}A_{(0)i}\right), \nn \\
\< {\cal O}^{\rm eq}_{1} \>_d &=& - \hat P\left(1+2\s u_{y}^{2}\right), \nn \\
\< {\cal O}^{\rm eq}_{2} \>_d\d_{ab} &=& - \hat P\left(\d_{ab}+2\s u_au_b\right). \nn
\eea
Using \eqref{u_red}, these become
\bea
\label{dualop3}
\< T^{\rm eq}_{ij} \>_d &=& \hat P \left(\eta_{ij} + 2\s\cosh^2\w\hat u_i\hat u_j\right),  \\
\< J_{i}^{\rm eq} \>_d  &=& 2\s\sinh\w\cosh\w\hat P\hat u_i\,,  \\
\< {\cal O}^{\rm eq}_{1} \>_d &=& - \hat P\left(1+2\s \sinh^{2}\w\right),  \\
\< {\cal O}^{\rm eq}_{2} \>_d &=& - \hat P\,,\\
\< {\cal O}^{\rm eq}_{\psi} \>_d &=& - \frac{2\sinh^2\w\hat P}{2\s-d}\,,\\
\< {\cal O}^{\rm eq}_{\z} \>_d &=& - \frac{\hat P}{2\s-d}\left(2\s\cosh^2\w-d\right)\,,
\eea
so that the equilibrium quantities are
\be
	\hat P =\frac{L}{L_{AdS}}P\,,\qquad \hat \e = \left(2\s\cosh^2\w-1\right)\hat P\,,\qquad \hat q=2\s\sinh\w\cosh\w\hat P\,.
\ee
Inserting the value of the pressure density \eqref{PressureAdSbbrane} for the AdS black brane allows to recover the correct reduced pressure, energy and charge density \eqref{thermo} as well as the dual operators \eqref{dualop} and \eqref{dualop1}.

Let us now discuss the dissipative part. We simply need to insert
$u^{\mu} = (u^i, 0, u^y)$ in $\<T_{\mu \nu}^{\rm diss}\>_{2\s}$ and reduce to $d$ dimensions.
The Landau-Lifshitz frame condition (\ref{LLcons}) becomes in the reduced theory
\bea
\label{LLconsd}
\hat{u}^i \< J_i^{\rm diss} \>_d &=& \tanh\w \< \cO_1^{\rm diss} \>_d  \nono \\
\hat{u}^i \< {T}_{ij}^{\rm diss} \>_d &=& -\tanh\w \< J_j^{\rm diss}  \>_d
\eea
and in particular one finds out that the reduced frame is \emph{not} in the Landau frame, so that some care is needed to extract the transport coefficients: we will use the frame independent
formulation discussed in \cite{Bhattacharya:2011tr}. The authors derive them without assuming any choice of frame but using invariance and symmetry considerations. This approach is thus well-suited to our case, since upon reduction one does not end naturally in the Landau or Eckhart frame. The method relies on ensuring that the divergence of the entropy current is positive semi-definite. In $2\s+1$ dimensions, the entropy current expressed in the Landau frame is
\be \label{EntropyCurrent}
	\<J^\mu_s\>_{2\s}=su^\mu\,,
\ee
and it obeys the divergence equation (see for example \cite{Bhattacharya:2011tr}),
\be \label{DivEntropyCurrent}
	\nabla_\mu\<J^\mu_s\>_{2\s} = -\nabla_\mu\left(\frac{u_\nu}T\right)\<T^{\mu\nu}_{diss}\>_{2\s}=-\frac1T\s_{\mu\nu}\<T^{\mu\nu}_{diss}\>_{2\s}\,.
\ee
For this to be positive semi-definite  
the shear viscosity, $\eta_{2\s}$,
\be
{P}_\mu^\k {P}^\l_\nu \<{T}^{\rm diss}_{\k\l}\>_{2\s}
- \frac{1}{2\s-1} {P}_{\mu\nu} {P}^{\k\l} \<T^{\rm diss}_{\k\l}\>_{2\s}
= - 2 \eta_{2 \s} {\s}_{\mu\nu}\,.  \label{TensorInv1}\\
\ee
must be non-negative, $\eta_{2\s} \geq 0$.

For charged fluids, the entropy current is given by \cite{Bhattacharya:2011tr}
\footnote{Note that our conventions relate to those of \cite{Bhattacharya:2011tr} by changing $\hat A_{(0)i}\to-\hat A_{(0)i}$ and consequently $F^{ij}_{(0)}\to -F^{ij}_{(0)}$. This has no impact on \eqref{TensorInv} or \eqref{ScalarInv}, but changes the relative signs in \eqref{VectorInv} as well as in the conservation equation for the reduced boundary stress-energy tensor \eqref{ConsEqT}.}
\be \label{ChargedEntropyCurrent}
	\<J^i_s\>_{d}=\hat s\hat u^i-\frac{\hat u_j}{\hat T}\<T^{ij}_{diss}\>_d-\frac{\hat \mu}{\hat T}\<J^i_{diss}\>_d\,.
\ee
Imposing the reduced Landau frame conditions \eqref{LLconsd}, 
we find that this coincides with the reduction of the entropy 
current \eqref{EntropyCurrent},
\be \label{ReducedEntropyCurrent}
	\<J^i_s\>_{d}=\hat s \hat u^i,\qquad \hat s = \frac{L\cosh\w}{L_{AdS}}s\,,
\ee
while the reduction of the divergence equation \eqref{DivEntropyCurrent} yields
\be
	\label{ReducedDivEntropyCurrent}
	\partial_i\<J^i_s\>_{d} = -\partial_i\left(\frac{\hat u_j}{\hat T}\right)\<T^{ij}_{diss}\>_{d} - \left[\partial_i\left(\frac{\hat\mu}{\hat T}\right)-\frac{\hat u_k}{\hat T}F_{(0)i}^k\right]\<J^i_{diss}\>_d\,,
\ee
which coincides with equation (2.19) of \cite{Bhattacharya:2011tr}.

The reduced shear viscosity, $\hat \eta$, the heat conductivity, $\hat \kappa$ and the bulk viscosity $\hat \zeta_s$
can then be extracted from the formul\ae:
\bea
\hat{P}^i_k \hat{P}^j_l \<{T}^{\rm diss}_{ij}\>_d
- \frac{1}{d -1} \hat{P}_{kl} \hat{P}^{ij} \<T^{\rm diss}_{ij}\>_d
&=& - 2 \hat \eta \hat{\s}_{kl} \,, \label{TensorInv}\\
\hat{P}^j_i \left(\<J^{\rm diss}_j\>_d + \frac{\hat q}{\hat \e + \hat P} \hat{u}^i \< {T}_{ij}^{\rm diss} \>_d \right) &=& -\hat \kappa \left(\hat{P}_{ij} \partial^j \frac{\hat \mu}{\hat T} + \frac{F_{(0)ij} \hat{u}^j}{\hat T}\right), \label{VectorInv}\\
\frac{\hat{P}^{ij} \<{T}^{\rm diss}_{ij}\>_d}{d -1} - \frac{\partial \hat P}{\partial \hat \e} \hat{u}^i \hat{u}^j \< {T}_{ij}^{\rm diss} \>_d +
\frac{\partial \hat P}{\partial \hat q} \hat{u}^i \< J_i^{\rm diss} \>_d &=&
- \hat \zeta_s \partial_i \hat{u}^i\,. \label{ScalarInv}
\eea
Using (\ref{LLconsd}) the last two become
\bea
\hat{P}^{ij} \<J^{\rm diss}_j\>_d \left(1 - \frac{\hat q}{\hat \e + \hat P} \tanh \w \right) &=& -\hat \kappa \left(\hat{P}^{ij} \partial_j \frac{\hat \mu}{\hat T} + \frac{F^{ij}_{(0)} \hat{u}_j}{\hat T}\right), \\
\frac{\hat{P}^{ij} \<{T}^{\rm diss}_{ij}\>_d}{d -1} +
\left(\frac{\partial \hat P}{\partial \hat \e} \tanh^2 \w  +
\frac{\partial \hat P}{\partial \hat q} \tanh \w\right) \< \cO_1^{\rm diss} \>_d
&=& - \hat \zeta_s \partial_i \hat{u}^i\,.
\eea
Using the conservation equations for the fluid: 
\bea
    \partial_i \<T^{ij}\>_d&=& F^{ij}_{(0)} \<J_i\>_d\,,\\
    \partial^i \<J_i\>_d&=&0\,,
\eea
yields:
\bea
     \partial_j \log m &=&  \frac{\cosh\omega}{\sinh\omega}\,\hat u\cdot\partial\w~\hat u_j  -\cosh^2\w~ \hat u\cdot\partial\hat u_j+\sinh\w\cosh\w~\hat u^iF_{(0)ij}\,,\\
    \hat u\cdot\partial\omega&=& \frac{\sinh  \omega~\cosh\w}{2(\s-1)\cosh^2\omega+1}\partial\cdot\hat u\,.
\eea
We also calculate 
\bea
 \<T^{diss}_{ij}\>_d&=&-2\eta_d \left[\cosh\w\hat\sigma_{ij}+\sinh\w \hat u_{(i}\left(\partial_{j)}\w+\frac12\sinh2\w\hat u\cdot\partial\hat u_{j)}-\cosh^2\w\hat u_kF^k_{(0)j)}\right)+\right.\nn\\
		&&\left.+\cosh\w\frac{\hat P_{ij}}{d-1}\left(1-\frac{(d-1)\cosh^2\w}{2(\s-1)\cosh^2\w+1}\right)\partial\cdot\hat u\right],\\
 \<J^{\rm diss}_j\>_d&=&\eta_d\cosh\w\left[\hat u_j\hat u\cdot\partial\w-\partial_j\w-\sinh\w\cosh\w\hat u\cdot\partial\hat u_j-\cosh^2\w\hat u_iF_{(0)j}^i\right],\\
 {\< \cO_{1}^{\rm diss} \>_d} &=&{\< \cO_{2}^{\rm diss} \>_d} ={\< \cO_{\psi}^{\rm diss} \>_d} =2\eta_d\frac{\cosh^2\w}{\sinh\w}\hat u\cdot\partial\w\,,\\
 {\< \cO_{\zeta}^{\rm diss} \>_d} &=&0\,,
\eea
so that finally
\bea
    \hat\eta&=&\eta_d\cosh\omega = L m^{2\s-1}\cosh\omega\,, \label{ShearViscosity}\\
    \hat\kappa&=&\frac{\eta_d\hat T}{\cosh\omega} = \frac{\s Lm^{2\s}}{2\pi\cosh^2\omega}\,,\label{HeatConductivity}\\
    \hat\z_s &=&\frac{2\eta_d\cosh\omega}{2\s-1}\left[\frac{2\s-d}{d-1}-\frac{2\sinh^2\omega\left((\s-1)\cosh^2\omega+\s\right)}{\left(2(\s-1)\cosh^2\omega+1\right)^2}\right].\label{BulkViscosity}
\eea
$\eta_d$ is the shear viscosity of the (reduced) neutral case
\be \label{ShearNeutral}
	\eta_d = \frac{L}{L_{AdS}} \eta_{2 \s}= L m^{2\s-1}\,,
\ee
where the first equality comes from the reduction, while in the second
equality we used the universal value of $\eta_{2 \s}$ for conformal, 
AdS black branes, (\ref{PressureAdSbbrane}).

Note that the transport coefficients 
(\ref{ShearViscosity})-(\ref{HeatConductivity})-(\ref{BulkViscosity}) 
are the universal
coefficients valid for any solution with the
same asymptotics as the black brane solution discussed in the
previous section. Using
\eqref{entropyReduced} and \eqref{ShearViscosity}, we find that
$\hat\eta/\hat s=1/4 \pi$ and the KSS bound \cite{KSSbound} is
saturated for charged branes,
as a consequence of the fact that it is saturated for the
conformal branes, similar to the neutral case \cite{Kanitscheider:2009as}.
This is a dynamical statement: the value of $\eta/s$ is fixed by the
requirements of regularity in the interior (singular solutions can have
$\eta/s$ smaller or larger than $1/4 \pi$).

We now turn to the ratio $\hat \zeta_s/\hat \eta$. As discussed in \cite{Kanitscheider:2009as},
in this ratio the factor $\eta_{2 \s}$ drops out and
the value of the ratio is fixed kinematically by the reduction:
any solution with the given asymptotics, regular or singular,
will have the same ratio. The same comment applies to 
the ratio $\hat \kappa/\hat \eta$. 
Our result for $\hat \zeta_s/\hat \eta$ can be compared
with a recent formula in \cite{Eling:2011ms}:
\be
    \frac{\hat\z_s}{\hat\h} = \sum_i\left(\hat s\frac{\ud\phi_i^h}{\ud\hat s}+\hat q_a\frac{\ud\phi_i^h}{\ud\hat q_a}\right)^2,
    \label{ElingOzFormula}
\ee
where $\hat q_a$ are conserved charge densities and $\phi^h_i$ are a
collection of scalar fields, evaluated at the event horizon,
and the formula
is valid in the Einstein frame where the entropy density $s$ is
given by the quarter of the horizon area. This formula reproduces
all known results and we
would like to check it against our result \eqref{BulkViscosity}.

The entropy and charge density in the Einstein frame are still given by \eqref{entropyReduced} and \eqref{thermo}
from which it is straightforward  to derive
\bea
    \ud(\log\hat  s)|_{\hat q} &=& -\frac{2(\s-1)\cosh^2\omega +1}{2\s\cosh\omega~\sinh\omega}~\ud\omega \nonumber \\
   \ud(\log\hat q)|_{\hat s} &=& \frac{2(\s-1)\cosh^2\omega +1}{(2\s-1)\cosh\omega~\sinh\omega}~\ud\omega \nonumber \\
          \ud( \psi_h)|_{\hat s} &=& \sqrt{\frac{2(d-1)}{(2\s-1)(2\s-d)}}\tanh\omega~\ud\omega \nonumber \\
          \ud( \psi_h)|_{\hat q} &=&-\sqrt{\frac{2(2\s-1)}{(d-1)(2\s-d)}}\frac{2(\s-d)\cosh^2\omega~+d}{2\s\cosh\omega~\sinh\omega}~\ud\omega \nonumber \\
          \ud( \z_h)|_{\hat s} &=&d(\z_h)|_{\hat q} = \sqrt{\frac{2(2\s-d-1)}{(2\s-d)}}\tanh\omega~\ud\omega
\eea
so that grouping everything together in \eqref{ElingOzFormula}, one does
recover \eqref{BulkViscosity}. This constitutes a very non-trivial check, since the two methods are completely different.

Moreover, direct computation yields 
\be
\frac{\hat\zeta_s}{\hat\eta} = 2\left(\frac1{d-1}
-\hat c_{s}^2\right) -\frac{4 \sinh^2\omega\left((\s-1)\cosh^2\omega+1\right)}{\left(2(\s-1)\cosh^2\omega+1\right)^2}
\ee
so that the bound proposed in \cite{Buchel:2007mf}
\be \label{bound}
    \frac{\hat\zeta_s}{\hat\eta} \geq  2\left(\frac1{d-1}-\hat c_{s}^2\right)
\ee
is always violated\footnote{See \cite{Buchel:2011uj}
for recent work containing other such examples.}, except if
\be
    \s<\hat\mu^2.
\ee
This is possible only if $\s <1$ since $\hat\mu^2 = \tanh^2\w \leq 1$
but for all values in (\ref{sigma_value}) $\s >1$. The equality
is achieved when either $\hat\mu=0$ (neutral case) or else $\hat\mu^2=\sigma$.
Let us emphasize again that the ratio $\hat\zeta_s/\hat\eta$ is fixed kinematically,
given asymptotics, so there is no reason to expect that a general system
would satisfy such an inequality. In the charged case the asymptotic behavior is
different from the neutral one since the presence of a chemical potential (and regularity at the horizon)
implies that a non-normalizable mode for the gauge field is turned on,
see (\ref{chemical}).

We note, however, that there is a similar looking
inequality to (\ref{bound}) that is saturated by the neutral branes and
is satisfied by the charged ones, namely (\ref{bound}) but with the adiabatic speed of
sound $\hat c_s^2$ replaced by $\hat c_q^2$
\be
\hat c_q^2 \equiv \left.\frac{\partial \hat P}{\partial \hat\epsilon}\right|_{\hat q}  = \frac{\cosh 2 \omega}{(2 \sigma-2) \cosh^2 \w +1}\,.
\ee
$\hat c_q^2$ reduces to the speed of sound of the conformal branes when $\w =0$ and furthermore
\be
    \frac{\hat\zeta_s}{\hat\eta} - 2\left(\frac1{d-1}
-\hat c_{q}^2\right) = \frac{ (\s-1) \sinh^2(2\w) }{ (2(\s-1) \cosh^2\w + 1)^2 }.
\ee
The right hand side is manifestly positive when $\s >1$, which is true for all values in (\ref{sigma_value}).
It would be interesting to check whether there are any counterexamples to this inequality.

The DC conductivity can be deduced using the Franz-Wiedemann law:
\be \label{DCConductivity}
    \hat\sigma_{DC}=\frac{\hat\k}{\hat T} =  \frac{\eta_d}{\cosh\w}=\frac{Lm^{2\s-1}}{\cosh\w}\,.
\ee
In order to make comparisons with other results easier, one can reexpress all the transport coefficients for the reduced AdS black brane in terms of the temperature and chemical potential:
\bea
    \hat\eta&=& L\left(\frac{2\pi\hat T}{\sigma}\right)^{2\s-1}\left(1-\hat\mu^2\right)^{-\sigma}\,, \label{ShearViscosity2}\\
    \hat\kappa&=&\frac{\s L}{2\pi}\left(\frac{2\pi\hat T}{\sigma}\right)^{2\s}\left(1-\hat\mu^2\right)^{1-\sigma}\,,\label{HeatConductivity2}\\
    \hat\z_s &=&\frac{2(2\s-d)\hat\eta}{(d-1)(2\s-1)}\left[1-\frac{2(d-1)\hat\mu^2\left(2\s-1-\s\hat\mu^2\right)}{(2\s-d)\left(2\s-1-\hat\m^2\right)^2}\right],\label{BulkViscosity2}\\
    \hat\sigma_{DC}&=&  L\left(\frac{2\pi\hat T}\sigma\right)^{2\s-1}\left(1-\hat\mu^2\right)^{1-\sigma}\,. \label{DCConductivity2}
\eea
Note that taking the neutral limit $\hat\mu\to 0$ in the DC conductivity
gives a finite contribution: indeed
our computation
represents the microscopic
fluctuations around the background, whether the latter is neutral or
charged.
The result for the conductivity can be compared with results from the
flavour branes approach,
\cite{Karch:2007pd,Hartnoll:2009ns,Charmousis:2010zz,Gouteraux:2011ce},
but only in the zero density limit. Indeed, \eqref{DCConductivity2} is
obtained by working out the fluctuations of the metric and gauge field
around a charged black hole, while in the case of flavour branes, the
gauge field lives on the brane in a neutral background and does not
backreact\footnote{We wish to thank E. Kiritsis and F. Nitti for discussions
on this point.}.

\section{Discussion and conclusions}

In this paper we set up holography for non-asymptotically AdS solutions
of a class of Einstein-Maxwell-Dilaton theories. This was achieved by
showing that these theories are related to AdS-Maxwell theory in
higher dimensions by means of a generalized dimensional reduction over
compact Einstein manifolds. `Generalized' here refers to the continuation
of the dimension of the compact space to non-integral values.
Such a generalized dimensional reduction was introduced in
\cite{Kanitscheider:2009as} and here we include
gauge fields and additional scalar fields in the analysis.

The relation to higher dimensional
AdS gravity controls both the UV and the IR behavior of the
strongly coupled dual QFT. The UV behavior is dictated by
a fixed point at $d+\epsilon$ dimensions,
where $\epsilon$ is the dimension of the compact space,
whose existence follows from the fact that the solution uplifts to an
asymptotically AdS solution. From the $d$-dimensional perspective
this translates into specific running of coupling constants.
The IR behavior near thermal equilibrium, the hydrodynamic
regime, is also controlled by the higher-dimensional theory.
The universal hydrodynamic behavior of AdS gravity implies
via dimensional reduction a specific
hydrodynamic behavior of EMD theories. In particular,
an entropy current with non-negative divergence in AdS 
reduces to a entropy current with the same property in the reduced
theory and
the transport coefficients are directly related to those
of AdS gravity \cite{Kanitscheider:2009as}. 

This leads to certain kinematical relations among 
the transport coefficients.
For example, the ratios of the bulk to shear viscosity
and conductivity to shear viscosity are fixed to specific 
values irrespectively of whether the bulk solution is regular or
singular in the interior. Furthermore, when there is a chemical
potential the putative bound on the bulk to shear viscosity 
proposed in \cite{Buchel:2007mf} is generically violated. 

The duality described here is not in general valid at all
energy scales. A prototype example for the dualities we discuss
is the holographic duality for non-conformal branes.
In that case, as discussed in detail in section 2
of \cite{Kanitscheider:2008kd}, one assumes that the effective
't Hooft coupling $g_{eff}^2N$ is fixed while $N^2$ is taken to infinity.
However, in these theories the effective coupling constant depends on
the energy scale so there is always a regime where $g_{eff}^2N$
grows faster than $N^2$ implying that the dilaton blows up and
a new description is needed (which for the case of D$p$ branes
is typically that of an M-brane).
Our current discussion is not tied to any specific dual theory
but we expect the same to be true here: the holographic description
would be valid below a certain energy scale.

The recent interest in these theories originates from the desire to
model holographically interesting IR fixed points, mostly having in
mind applications to condensed matter systems. Models that interpolate
between the IR behavior described here
and an AdS region in the UV have been considered,
for example, in \cite{Goldstein:2009cv,Charmousis:2010zz,Perlmutter:2010qu,Goldstein:2010aw,Bertoldi:2010ca,Bertoldi:2011zr,Gouteraux:2011ce}. One would expect on general grounds to
be able to model the IR region without a reference to such UV completion and
indeed our discussion provides precisely such a description.

There are many possible extensions and generalizations of this work.
In section \ref{section:oxi} we described an array of theories
which are linked with AdS-Maxwell gravity in higher dimensions
but we only worked out the holographic dictionary and the
hydrodynamic regime for one of them. It would be interesting
to work out the holographic data for the entire class. For example,
the case of two gauge fields is interesting since such systems
could be used to describe holographically imbalanced
superconductors, see \cite{Erdmenger:2011hp} for recent work in this
direction. The case where the higher-dimensional theory is AdS-Gauss-Bonnet is under investigation, \cite{Charmousis:2011}.
More generally, it would be interesting to map out all possibilities
where such a generalized dimensional reduction could be used in order
to set up holography.

\acknowledgments
We would like to thank C. Charmousis, E. Kiritsis and F. Nitti for discussions.
This work is part of the research program of the `Stichting voor
Fundamenteel Onderzoek der Materie' (FOM), which is financially
supported by the `Nederlandse Organisatie voor Wetenschappelijk
Onderzoek' (NWO). JS, MS and KS acknowledge support via an NWO
Vici grant. BG wishes to thank ITFA for hospitality
at various stages of this project.

\appendix

\section{Appendix\label{Appendix}}

\subsection{Diagonal reduction of $(2\sig+1)$-AdS-Maxwell theories
\label{Appendix:DiagonalKK}}

Our starting point is the Einstein-AdS action with a cosmological constant in $2\sig+1$ dimensions \eqref{EinsteinAdSAction} and a Maxwell field strength:
\be
    S_{(2\sigma+1)}=\frac1{16\pi G_N^{(2\s+1)}}\int\ud^{2\sigma+1}\sqrt{-g_{(2\sigma+1)}}\left[R_{(2\s+1)}-\frac14F^2-2\Lambda\right].
\ee
It has equations of motion
\bea
    G_{AB}+\Lambda g_{AB}&=&\frac12F_{AC}F_B^{\phantom{1}C}-\frac18F^2g_{AB},\label{EeqA1}\\
    \nabla_AF^{AB}&=&0\,.\label{MaxwellEqA1}
\eea
We wish to perform a reduction to an Einstein-Maxwell-Dilaton theory with a static Ansatz:
\be
    \ud s^2_{(2\sigma+1)}=e^{2\a\phi}\ud s^2_{(d+1)} + e^{2\b\phi}\ud X^2_{(2\sigma-d)}, \qquad A_A=(A_M(x^N),A_a=0)
    \label{KKStatic1}
\ee
where $\ud X^2_{(2\sigma-d)}$ is the metric of a $(2\sigma-d)$-dimensional Einstein space, \eqref{EinsteinSpace}, with normalised curvature $\lambda_{(2\sig-d)}$.

For a diagonal Ansatz, it is consistent to take all scalar fields along each reduced direction equal, see \cite{Lu:1995yn} for the generic (toroidal) case. Nonetheless, let us check that such an Ansatz is consistent by reducing Einstein's equations directly and writing out the action from which they derive.

Using the tetrad formalism, the higher-dimensional Einstein tensor $G^{(2\s+1)}_{AB}$ can be projected on the external and internal coordinates:
\bea
    G^{(2\s+1)}_{MN}&=&G^{(d+1)}_{MN}+\left[(d-1)\a^2+(2\s-d)2\a\b-(2\s-d)\b^2\right]\partial_M\phi\partial_N\phi-\nonumber\\
    &&-\left[(d-1)\a+(2\s-d)\b\right]\nabla_M\nabla_N\phi-\nonumber\\
    &&-\frac12g_{MN}^{(d+1)}\left\{R_{(2\s-d)}e^{2(\a-\b)\phi}-2\left[(d-1)\a+(2\s-d)\b\right]\square\phi-\right.\\
    &&\left.-\left[(d-1)(d-2)\a^2+2(2\s-d)(d-2)\a\b+(2\s-d)(2\s-d+1)\b^2\right]\partial\phi^2\right\}\label{GmunuA}\nonumber\\
    G^{(2\s+1)}_{ab}&=&G^{(2\s-d)}_{ab}-\frac12g_{ab}^{(2\s-d)}e^{2(\b-\a)\phi}\left\{R_{(d+1)}-2\left[d\a+(2\s-d-1)\right]\square\phi-\right.\\
    &&\left.-\left[d(d-1)\a^2+2(d-1)(2\s-d-1)\a\b+(2\s-d)(2\s-d-1)\b^2\right]\partial\phi^2\right\}\label{GmnA}\nonumber
\eea
where $G^{(d+1)}_{MN}$ and $G^{(2\s-d)}_{ab}$ are respectively the Einstein tensor associated to the $(d+1)$-dimensional metric and to the $(2\s-d)$-dimensional compact space $X_{(2\s-d)}$.
Then, taking the trace of $G_{AB}^{(2\s+1)}$, one finds the Ricci scalar
\bea
    R_{(2\sigma+1)}e^{2\a\phi} &=& R_{(d+1)}+e^{2(\a-\b)\phi}X_{(2\sig-d)}-2(d\a+(2\sig-d)\b)\Box\phi-\\
    &&-\left[d(d-1)\a^2+(2\sig-d)(2\sig-d+1)\b^2+2(2\sig-d)(d-1)\a\b\right]\partial\phi^2\,,\nn
\eea
while
\be
    \det g_{(2\sig+1)}=e^{\left[2(d+1)a+(2\sigma-d)b\right]\phi}\det g_{(d+1)}.
\ee
Setting the overall conformal factor in $\phi$ in the action to unity\footnote{\emph{e.g.}, going to the Einstein frame.} requires
\be
    (2\sig-d)\b=(1-d)\a
\ee
upon which
\be
    R_{(2\sigma+1)}e^{2\a\phi} = R_{(d+1)}-2\a\Box\phi-(d-1)\frac{2\sig-1}{2\sig-d}\a^2\partial\phi^2+e^{2\frac{2\sig-1}{2\sig-d}\a\phi}R_{(2\sig-d)}\,.
\ee
In order to have a canonically normalised kinetic term for the scalar, we then set
\be
    \a=-\sqrt{\frac{(2\sig-d)}{2(d-1)(2\sig-1)}}=-\frac\d2\quad \Leftrightarrow \quad \d=\sqrt{\frac{2(2\sig-d)}{(d-1)(2\sig-1)}}
\ee
so that the bulk action naively becomes
\bea
    S_{(d+1)}&=&\frac{1}{16\pi G_N^{(d+1)}}\int_{\mathcal M}\ud^{d+1}x\sqrt{-g_{(d+1)}}\left[R_{(d+1)}-\half\partial\phi^2-\frac14e^{\d\phi}F^2-2\Lambda e^{-\d\phi}+\right.\nn\\
    && \qquad\left.+R_{(2\sigma-d)}e^{-\frac{2\phi}{(d-1)\d}}\right]-\frac{1}{16\pi G_N^{(d+1)}}\int_{\partial\mathcal M}\ud^{d}x\sqrt{-h_{(d)}}\,\d\, n^M\partial_M\phi\,.
    \label{AdSMaxD+1}
\eea

To check that this is correct, we can also replace in \eqref{GmunuA} and \eqref{GmnA}
\bea
    G^{(2\s+1)}_{MN}&=&G^{(d+1)}_{MN}-\frac12\partial_M\phi\partial_N\phi-\frac12g_{MN}^{(d+1)}\left[R_{(2\s-d)}e^{-\frac{2\phi}{(d-1)\d}}-\frac12\partial\phi^2\right]\label{GmunuA2}\\
    G^{(2\s+1)}_{ab}&=&G^{(2\s-d)}_{ab}-\frac12g_{ab}^{(2\s-d)}e^{\frac{2\phi}{(d-1)\d}}\left[R_{(d+1)}+\frac{2}{(d-1)\da}\square\phi-\frac12\partial\phi^2\right].\label{GmnA2}
\eea
and reexpress Einstein's equations \eqref{EeqA1}:
\bea    G^{(d+1)}_{MN}&=&\frac12\partial_M\phi\partial_N\phi+\frac{e^{\d\phi}}2F_{MP}F^{\phantom{1}P}_M+\nonumber\\
                &&+\frac{g_{MN}^{(d+1)}}2\left[R_{(2\s-d)}e^{-\frac{2\phi}{(d-1)\d}}-\frac12\partial\phi^2-\frac{e^{\d\phi}}4F^2-2\Lambda e^{-\d\phi}\right]\label{GmunuA3}\\
    G^{(2\s-d)}_{ab}&=&\frac{g_{ab}^{(2\s-d)}}2e^{\frac{2\phi}{(d-1)\d}}\left[R_{(d+1)}+\frac{2\square\phi}{(d-1)\d}-\frac12\partial\phi^2-\frac{e^{\d\phi}}4F^2-2\Lambda e^{-\d\phi}\right]\label{GmnA3}
\eea
In \eqref{GmunuA3}, we recognise the lower-dimensional equation of motion for the metric, as derived from \eqref{AdSMaxD+1}. Taking the trace of \eqref{GmnA3} and replacing again in \eqref{GmnA3}, one finds that $\mathbf X_{(2\s-d)}$ must be an Einstein space, that is
\be
R^{(2\s-d)}_{ab}=\frac{R^{(2\s-d)}}{2\s-d}g^{(2\s-d)}_{ab}.
\ee
The lower-dimensional Ricci scalar can be derived from \eqref{GmunuA3} or  \eqref{GmnA3}:
\bea
    R_{(d+1)}&=&\frac12\partial\phi^2+\frac{(d-3)e^{\da\phi}}{4(d-1)}F^2+\frac{d+1}{d-1}2\Lambda e^{-\d\phi}-\frac{d+1}{d-1}R_{(2\s-d)}e^{-\frac{2\phi}{(d-1)\d}}\\
    R_{(d+1)}&=&\frac12\partial\phi^2+\frac{e^{\da\phi}}4F^2+2\Lambda e^{-\d\phi}-\frac{2\s-d-2}{2\s-d}R_{(2\s-d)}e^{-\frac{2\phi}{(d-1)\d}}-\frac{2\square\phi}{(d-1)\d}\,.
\eea
Subtracting the two previous equations yields the dilaton equation of motion:
\be
    \square\phi=\frac{\da}4e^{\da\phi}F^2-2\d\Lambda e^{-\d\phi}+\frac{2}{(d-1)\d}R_{(2\s-d)}e^{-\frac{2\phi}{(d-1)\d}},
\ee
identical to that derived from \eqref{AdSMaxD+1}, while the other combination gives back the trace of Einstein's equations. Finally, the lower-dimensional Maxwell equation follows straightforwardly from the higher-dimensional one \eqref{MaxwellEqA1}.

The metric Ansatz becomes
\be
    \ud s^2_{(2\sigma+1)}=e^{-\d\phi}\ud s^2_{(d+1)} + e^{\frac\phi\d\left(\frac{2}{d-1}-\d^2\right)}\ud X^2_{(2\sigma-d)}.
    \label{KKStatic}
\ee
We have also defined the lower-dimensional Newton's constant $G_N^{(d+1)}= G_N^{(2\sig+1)}/V_{(2\sig-d)}$, where $V_{(2\sig-d)}$ is the volume of $\mathbf X^{(2\sig-d)}$. The term in $\Box\phi$ does not impact the lower-dimensional equations generates a boundary term on $\partial\mathcal M$, the boundary of the bulk spacetime $\mathcal M$ defined by its unit normal vector $n^\mu$ and boundary metric $h_{(d)}$. It has no impact on the equations of motion, but would be important for the computation of the Euclidean action on-shell.

Let us also consider the reduction of the Gibbons-Hawking-York boundary term, which involves the trace $K_{(2\sig)}$ of the extrinsic curvature of spacetime. Using the Ansatz \eqref{KKStatic}, it is a matter of calculation to show that
\be
    \sqrt{-h_{(2\sig)}}=e^{-\frac\d2\phi}\sqrt{-h_{(d)}}\,,\qquad K_{(2\sig)}=e^{\frac\d2\phi}\left[K_{(d)}-\frac\d2n^M\partial_M\phi\right].\label{KKreducExtrinsicCurvature}
\ee

We can move on and deal with the Gibbons-Hawking-York boundary term:
\bea
    S^{GBH}_{(2\sig)}&=&-\frac{1}{8\pi G_N^{(2\sig+1)}}\int_{\partial\mathcal M} \sqrt{-h_{(2\sig)}} \,\ud^{2\sig}x\,K_{(2\sig)}\nn\\
                &=&-\frac{1}{8\pi G_N^{(d+1)}}\int_{\partial\mathcal M} \sqrt{-h_{(d)}} \,\ud^{2\sig}x\left[K_{(d)} -\frac\d2n^M\partial_M\phi\right],
\eea
where we have used \eqref{KKreducExtrinsicCurvature}. The first term is the lower-dimensional Gibbons-Hawking-York boundary term, while the second term is exactly the one needed so that the boundary term coming from the reduction of the higher-dimensional Ricci scalar is cancelled, see \eqref{AdSMaxD+1}. In the end, only the $(d+1)$-dimensional GHY term is left.

Let us now make contact with the generic Einstein-Dilaton action \eqref{ED2LiouvilleAction}
\bea
    S_{(d+1)}&=& \frac{1}{16\pi G_N^{(d+1)}}\int_{\mathcal M}  \ud^{d+1}x\sqrt{-g}\left[R-\frac12(\partial \phi)^2-\frac14 e^{\g\phi}F^2-2\Lambda_1 e^{-\delta_1\phi}-2\Lambda_2 e^{-\delta_2\phi}\right]-\nn\\
    &&\qquad\qquad\qquad-\frac{1}{8\pi G_N^{(d+1)}}\int_{\partial\mathcal M} \sqrt{-h_{(d)}} \,\ud^{2\sig}x\,K_{(d)}\,, \label{EMDAction}
\eea
As shown above, using the metric Ansatz:
\be
    \ud s^2_{(2\sigma+1)}=e^{-\d_1\phi}\ud s^2_{(d+1)} + e^{\frac\phi\d_1\left(\frac{2}{d-1}-\d_1^2\right)}\ud X^2_{(2\sigma-d)}\,,
    \label{KKStaticToroidal}
\ee
and setting
\be
    \Lambda_1=\Lambda\,,\qquad R_{(2\sigma-d)}=-2\Lambda_2\,, \qquad \delta_2 = \frac{2}{(d-1)\da_1}
\ee
so that the first Liouville potential in \eqref{ED2LiouvilleAction} originates from the higher-dimensional cosmological constant $\Lambda$ and the second one from the curvature of the internal space, this action is a consistent reduction of the Einstein-AdS-Maxwell action.
The exponent $\da_1$ is related to the number of reduced dimensions as:
\be
    \d_1=\sqrt{\frac{2(2\sig-d)}{(d-1)(2\sig-1)}} \Leftrightarrow 2\sig = \frac{2d-(d-1)\d_1^2}{2-(d-1)\d_1^2}
    \label{RelationDaSigTorStatic}
\ee
from which $2\s$ vary with $\d_1$ in the following way:
\begin{table}[!ht]
 \begin{tabular}{|c|lcr|lcr|lcr|}
 \hline
 $\d_1^2$ &$0^+$&            & $\left(\frac{2}{d-1}\right)^-$     & $\left(\frac{2}{d-1}\right)^+$    &          &$\left(\frac{2d}{d-1}\right)^-$&$\left(\frac{2d}{d-1}\right)^+$&          &$+\infty$\\
 \hline
         &     &            & $+\infty$                          &                                   &         &$0^-$&     &          &$(d+1)^-$\\
  $2\s+1$&     & $\nearrow$ &           &          &$\nearrow$&     &     &$\nearrow$&      \\
         &$(d+1)^+$&            &           & $-\infty$&          &     &$0^+$&          &\\
  \hline
 \end{tabular}
\end{table}

A consistent range of dimension values for the higher-dimensional theory is $0\leq\d_1^2\leq2/(d-1)$.

 To extend $\d_1$ to the complementary range, let us reverse the origins of the Liouville potentials in \eqref{EMDAction}, whereupon $\d_1$ has to be set to
 \be
    \d_1^2=\frac{2(2\sig-1)}{(d-1)(2\sig-d)} \Leftrightarrow 2\sig = \frac{d(d-1)\d_1^2-2}{(d-1)\d_1^2-2}\,,
    \label{RelationDaSigCurvedStatic}
\ee
with the $\Lambda_1$ Liouville now descending from the curvature of the internal space $R_{(2\s-d)}$, the $\Lambda_2$ one from the higher-dimensional constant $\Lambda$.
The range of dimension values spanned by $\da$ is now:
\begin{table}[!ht]
 \begin{tabular}{|c|lcr|lcr|lcr|}
 \hline
 $\d_1^2$ &$0^+$&            & $\left(\frac{2}{d-1}\right)^-$     & $\left(\frac{2}{d-1}\right)^+$    &          &$\left(\frac{2d}{d-1}\right)^-$&$\left(\frac{2d}{d-1}\right)^+$&          &$+\infty$\\
 \hline
           &$2^-$&            &         &$+\infty$&          &         &$(d+2)^-$&          &         \\
  $2\s+1$&     & $\searrow$ &         &         &$\searrow$&         &         &$\searrow$&         \\
           &     &            &$-\infty$&         &          &$(d+2)^+$&         &          &$(d+1)^+$\\
  \hline
 \end{tabular}
\end{table}

Its consistent restriction $\d_1^2>2/(d-1)$ is indeed the complementary of the previous one. Note that there is an upper bound on $\d_1$,
\be
    \d_1^2<\delta^2_{max}=\frac{2d}{d-1}
\ee
which reflects the fact that the space $\mathbf X_{(2\sigma-d)}$ only has non-zero curvature if $2\s>d+2$.

The metric Ansatz is:
\be
    \ud s^2_{(2\sigma+1)}=e^{-\frac{2\phi}{(d-1)\d_1}}\ud s^2_{(d+1)} + e^{\frac\phi\d_1\left(\d_1^2-\frac{2}{d-1}\right)}\ud X^2_{(2\sigma-d)}.
    \label{KKStatic2}
\ee
The higher-dimensional theory is still Einstein-AdS but the inclusion of a Maxwell field in the higher-dimensional action requires $\gamma=\frac{2}{(d-1)\d_1}$ in the lower-dimensional EMD action.

\subsection{Non-diagonal reduction of AdS theories along an $\mathbf S^1$\label{Appendix:NonDiagonalS1KK}}

Let us start again from the Einstein-AdS theory \eqref{EinsteinAdSAction}. We then reduce along a circle $\mathbf S^1$, this time by means of a non-diagonal Ansatz
\be
    \ud s_{(d+2)}^2 = e^{2\a\phi}\ud s^2_{(d+1)} + e^{-2(d-1)\a\phi}\left(\ud y+\mathcal A\right)^2,\qquad 2\sig=d+1\,.
\ee
with
\be
 \mathcal A=\mathcal A_{M}\ud x^M\,.
\ee
We use calligraphic notation to distinguish gauge fields arising in the $(d+1)$-dimensional theory through the compactification from those descending from higher-dimensional ones. Then,
\be
    e^{2\a\phi}R_{(d+2)}=R_{(d+1)}-d(d-1)\a^2\partial\phi^2-\frac14e^{-2d\a\phi}\mathcal F_{MN}\mathcal F^{MN}
\ee
discarding the $O(\Box\phi)$ term at this point. Normalising the kinetic term for the scalar field automatically gives
\be
    \a=-\frac1{\sqrt{2d(d-1)}}
\ee
and setting
\be
    \da=\sqrt{\frac{2}{d(d-1)}}\,, \quad \ga=\sqrt{\frac{2d}{d-1}}\,,\quad \ga\da=\frac2{d-1}
\ee
we recover the EMD action \eqref{EMD2LiouvilleAction}
\be
    S_{(d+1)}=\frac1{16\pi G_N^{(d+1)}}\int \ud^{d+1}x\sqrt{-g_{(d+1)}}\left[R_{(d+1)}-\half\partial\phi^2-\frac14 e^{\frac{2\phi}{(d-1)\da}}\mathcal F_{\mu\nu}\mathcal F^{MN}-2\Lambda e^{-\da\phi}\right],
\ee
albeit with a single Liouville potential: reducing over an $\mathbf S^1$ does not generate a potential due to its zero curvature.
The metric Ansatz becomes:
\be
    \ud s_{(d+2)}^2 =  e^{-\da\phi}\ud s^2_{(d+1)} + e^{\frac\phi\da\left(\frac2{d-1}-\da^2\right)}\left(\ud y+\mathcal A\right)^2.
\ee

\bibliographystyle{JHEP}

\end{document}